\documentclass[%
reprint,superscriptaddress,
 amsmath,amssymb,
 aps,
pra,
]{revtex4-2}

\usepackage{graphicx}
\usepackage{dcolumn}
\usepackage{bm}
\usepackage{siunitx}
\usepackage{array}
\usepackage{longtable}
\usepackage{booktabs}
\usepackage{hyperref}
\usepackage{comment} %
\usepackage{xcolor}
\usepackage{float}
\makeatletter
\def\maketitle{
\@author@finish
\title@column\titleblock@produce
\suppressfloats[t]}
\makeatother

\begin{document}

\title{Sn-InAs nanowire shadow-defined Josephson junctions}

\author{Amritesh Sharma}
\altaffiliation{These authors contributed equally to this work.}
\affiliation{Department of Physics and Astronomy, University of Pittsburgh, Pittsburgh, PA 15260, USA}

\author{An-Hsi Chen}
\altaffiliation{These authors contributed equally to this work.}
\affiliation{Univ. Grenoble Alpes, Grenoble INP, CNRS, Institut N\'eel, 38000 Grenoble, France}

\author{Connor P. Dempsey}
\affiliation{Electrical and Computer Engineering, University of California Santa Barbara, Santa Barbara, CA 93106, USA}

\author{Amrita Purkayastha}
\affiliation{Department of Physics and Astronomy, University of Pittsburgh, Pittsburgh, PA 15260, USA}

\author{Mihir Pendharkar}
\affiliation{Electrical and Computer Engineering, University of California Santa Barbara, Santa Barbara, CA 93106, USA}

\author{Susheng Tan}
\affiliation{Department of Electrical and Computer Engineering, and Petersen Institute of NanoScience and Engineering, University of Pittsburgh, PA 15260 USA}

\author{Christopher J. Palmstr{\o}m}
\affiliation{Electrical and Computer Engineering, University of California Santa Barbara, Santa Barbara, CA 93106, USA}
\affiliation{California NanoSystems Institute, University of California Santa Barbara, Santa Barbara, CA 93106, USA}
\affiliation{Materials Department, University of California Santa Barbara, Santa Barbara, CA 93106, USA}

\author{Sergey M. Frolov}
\affiliation{Department of Physics and Astronomy, University of Pittsburgh, Pittsburgh, PA 15260, USA}

\author{Moïra Hocevar$^\ast$}
 \email{moira.hocevar@neel.cnrs.fr}
\affiliation{Univ. Grenoble Alpes, Grenoble INP, CNRS, Institut N\'eel, 38000 Grenoble, France}%

\date{\today}

\begin{abstract}
Interest in hybrid electronic devices for quantum science is driving the research into superconductor-semiconductor materials combinations. Here we study InAs nanowires coated with shells of $\beta$-Sn. The wires grow via the vapor-liquid-solid mechanism out from (001) InAs substrates along two orientations, forming a criss-crossing landscape. This allows us to define nanowire-shadow junctions during the low temperature Sn shell deposition by carefully choosing the deposition angle. We find that the Sn shells are uniform in thickness and the grains have a preferential in-plane epitaxial relationship with InAs. The interface between Sn and InAs is abrupt and we do not observe interdiffusion. In our nanowire devices, Sn induces a superconducting gap of order 600 µeV, switching currents reaching values up to 500 nA, and critical magnetic fields along the nanowire of up to 1.3 T. These characteristics can be leveraged in the design of superconducting transmon qubits, parametric microwave amplifiers as well as for the investigation of triplet and topological superconductivity.

\end{abstract}

\maketitle

\section{\label{sec:level1} Introduction}


The rapid advance of solid-state quantum computing has so far been largely powered by well-studied materials such as aluminum and silicon. As the complexity of quantum processors increases, the question of whether the workhorse materials of the day will be able to carry the field to full-scale quantum computers gains in prominence. Hybrid superconductor-semiconductor materials were developed in parallel in an effort that was initially motivated by mesoscopic physics such as Majorana Zero Modes, which can be of relevance for topological quantum computing. Among the hybrid materials systems, quantum well or selective area-grown structures (SAG) offer forward-looking advantages such as scalability, while bottom-up grown nanowires or exfoliated flakes are best suited for rapid prototyping of materials combinations.
Hybrid materials were also used in proof-of-principle quantum devices such as transmon qubits and parametric amplifiers. The emphasis there has been on replacing a standard Al/$\mathrm{AlO_x}$ shadow Josephson junction with an electrically-tunable junction based on a nanowire \cite{LangeDiCarlo2015_Gatemon,Casparis2019_coupler,HaysDevoret2021_AndreevQubit,splitthoff2024gate}, a quantum well \cite{LarsenMarcus2015_Gatemon,hao2024kerr,phan2023gate,sagi_gate_2024}, or a van der Waals material \cite{butseraen2022gate, Oliver2018_Graphene, KCFong2021_vdWQubits, RVijay2022_GrapheneJunction}. 
The majority of these prototypes have still relied on Al as a superconductor. While offering reliable contacts to a variety of low-density materials as well as high microwave range quality factors, Al is also a relatively low transition temperature superconductor. An interesting question to explore is whether device performance can be improved or functionality enhanced by pivoting beyond aluminum. Other elemental superconductors such as Nb \cite{CarradJesper2020_AlNbTaV_InAs,GunelSchapers2012_NbInAs,PerlaShapers2021_NbInAs}, Pb \cite{Giazotto2015_PbInAs,SabbirKrogstrup2023_PbInSb}, V \cite{Giazotto2011_VInAs, Krogstrup2019_VInAs_noJJ}, Ta \cite{CarradJesper2020_AlNbTaV_InAs,vanschijndel2024_Ta} have been under investigation in hybrid devices.

Several efforts, including this one, have zeroed-in on Sn \cite{pendharkar2019paritypreserving, Aranya2023_SnInAsSAG, Po2022_SnSmashJJ,Po2024_SnSecondOrderJJ} which is a material that has two easily accessible crystalline phases, the $\alpha$-phase that is a semimetal, and a $\beta$-phase which is a superconductor with a critical temperature of 3.7 K, three times that of Al. In hybrid devices, Sn has demonstrated field-resilient superconductivity, with critical magnetic fields of several Tesla, hard-induced superconducting gap, as well as the coveted 2e-charging effect, previously only observed in Al \cite{shen_parity_2018}. 
Previous work showed that Sn layers grown on InSb nanowires formed in the tetragonal $\beta$-phase though the presence of the $\alpha$-phase in some grains could not be ruled out \cite{pendharkar2019paritypreserving}. The $\alpha$-phase is likely promoted by a close lattice match between $\alpha$-Sn and InSb. Furthermore, no reliable recipe for selectively etching Sn on group III-V semiconductors and retain superconducting device performance has been identified so far. These are significant limitations for exploring the full potential that Sn-based devices have to offer. 

To overcome these, here we use InAs nanowires produced using gold-catalyzed vapour-liquid-solid (VLS) growth on (001) substrates. On the one hand, InAs differs from InSb because it is not lattice matched to Sn in either of the phases, and this is expected to promote the $\beta$-phase of Sn. On the other hand, the nanowires on (001) substrates grow in two directions forming a criss-crossing landscape. This offers a straightforward pathway to create junctions in the Sn shell by criss-crossing nanowires shadowing each other without the need to chemically etch Sn.
We grow inclined and criss-crossing InAs nanowires optimized for dimensions, density and parallelism. 
In this, we identify the annealing temperature  of the substrate after gold colloid deposition as a key parameter. 
The Sn shells on InAs nanowires are uniform owing to low-temperature deposition, with evidence of preferential in plane epitaxial relation between Sn and InAs.
Upon hydrogen cleaning of the nanowire surface, we find the InAs/Sn interfaces to be abrupt with no evidence for interdiffusion. 
In low temperature measurements of Sn-InAs shadow junctions we find hard induced superconducting gap of approximately 600 $\mu$eV, comparable to the bulk gap of Sn. The maximum superconducting switching current reaches 0.5 $\mu$A. Superconducting properties persist to parallel magnetic fields in excess of 1.3 Tesla. 


The methods of shadow junction fabrication using (001) growth can be extended to test other superconductor-semiconductor combinations. In a parallel effort, these nanowires have been used to demonstrate superconducting gate-tunable transmon qubits with microsecond relaxation and coherence times, illustrating the potential of Sn-based quantum devices. High critical currents can also be used to develop single-nanowire based parametric quantum amplifiers.

\section{Experimental Methods}

\subsection{Background: Hybrid Structures Growth}



The deposition of thin superconductor metal films can be performed with basic techniques such as room temperature thermal evaporation combined with lithographically defined metal leads \cite{Delsing2014_InAs, GunelSchapers2012_NbInAs, Giazotto2015_PbInAs}. 
Higher in-plane critical magnetic fields require thinner superconducting layer thicknesses.
Uniformity and crystallinity are greatly influenced by the surface and kinetics at various stages of growth. 
Monocrystalline films may require high temperatures as observed for V\cite{Moodera2016_V}. 
But higher temperatures risk chemical reactions with underlying InAs nanowire \cite{Krogstrup2019_VInAs_noJJ, Schapers2017_LowTemp}. 
For materials such as Nb that have low surface diffusion lengths at ambient temperatures, clustering can happen at initial growth stages leading to rougher films. 
Deposition at near-orthogonal angles to nanowire facets contributes to smoother films of Ta and Nb, yet results in an amorphous/nanocrystalline phase \cite{Schapers2017_LowTemp, CarradJesper2020_AlNbTaV_InAs}.
Moreover, when grown at room temperature, Ta tends to form its high-resistivity, metastable $\beta$-phase, as observed in \cite{vanschijndel2024_Ta}. 
For other materials like Al \cite{Schapers2017_LowTemp}, Pb \cite{Jesper2021_PbLowTemp} and Sn \cite{pendharkar2019paritypreserving, Sabbir2020_SnTransparentGatable,khan2023} it is possible to obtain smooth and epitaxial, although polycrystalline, films when growing at lower temperatures. 
Despite growth at low temperatures, dewetting can occur when bringing back substrates to room temperature and techniques like appropriate capping with another material is needed \cite{Jane2023_SnCappingLayer}. 

In hybrid structures, it is crucial to optimize the superconductor-semiconductor interface in terms of roughness, interdiffusion and crystallinity \cite{Krogstrup2015,pendharkar2019paritypreserving,Jesper2021_PbLowTemp}.
Simultaneously, the fabrication of quantum devices requires either removing or leaving a segment of a semiconductor free of the superconductor. Since removal risks damage to the semiconductor, shadow junctions such as those used here can offer avantages. 
The elaboration of $in-situ$ shadows requires a fine control of the nanowire growth orientation, and morphology, so that they cross and do not merge.
Using the vapor-liquid-solid mechanism (VLS) with a metallic catalyst in a molecular beam epitaxy (MBE) chamber, one can effectively control the growth direction of the nanowires by adjusting the growth temperature and the flux of group III and group V species, and using different substrate orientations \cite{Zeng2020,chen2016direct}.
The growth of InAs nanowires on (111)B InAs substrates is well studied and modeled \cite{Jesper2021_PbLowTemp, Krogstrup2015, dubrovskii2016length,froberg2007diameter,mosiiets2024dual}. 
The nucleation mechanisms and optimal growth conditions for inclined InAs nanowires grown on (001) InAs that we use here are less studied and should be explored deeper.

\subsection{Nanowire growth} 

Our InAs nanowires are synthesized using the gold-assisted VLS mechanism in a molecular beam epitaxy (MBE) reactor in Grenoble. The nanowires grow from gold colloids, of radius ranging from 10 to 30 nm, used as catalysts. They are spun over a substrate sample freshly deoxidized in a bath of $\mathrm{NH_{4}OH:H_{2}O}$ for 5 min. The samples are then mounted on molybdenum blocks and degassed in ultra high vacuum at 250$^{\circ}$C in the introduction chamber of the MBE. In the main MBE chamber, the temperature of the substrate is increased to an annealing temperature ranging between 440$^{\circ}$C and 560$^{\circ}$C during 1 min under an $\mathrm{As_4}$ beam equivalent pressure of $1e-5$ Torr and then cooled down to 420$^{\circ}$C. After stabilizing the temperature for 5 min, the growth is started by opening the shutter of the In effusion cell at an In beam equivalent pressure of $2.5e-7$ Torr. After 20-30 min of growth, the In shutter is closed and the sample is cooled down to 350$^{\circ}$C under As flux, and to room temperature in vacuum. 
Nanowire growth is monitored using a Reflection High-Energy Electron Diffraction (RHEED) setup. As the nanowires grow inclined on (001) oriented substrates, the RHEED pattern appears as chevrons \cite{Balakrishnan2006}.

InAs nanowires typically grow in the <111>B direction. On (001) oriented InAs substrates, the nanowires grow inclined along the two horizontal directions $[1\overline{1}0]$ at 35.2$^{\circ}$ angle with respect to the substrate surface. In this configuration, nanowires cross with or without touching each other. On average, every other wire is oriented in an opposite direction. This growth strategy allows nanowires to naturally shadow each other in subsequent steps. 

\subsection{Shadow Deposition of Sn} 

Metal deposition is performed ex-situ at cryogenic temperatures in an ultra high vacuum chamber at UCSB. At this point, the nanowire chips are exposed to the atmosphere and the nanowires require native oxide removal. The growth chips are attached to molybdenum blocks through gallium bonding and placed in a vacuum chamber where they undergo a process of atomic hydrogen cleaning at 380$^{\circ}$C (thermocouple temperature) for 30 minutes, under an operating pressure of $5e-6$ Torr, primarily composed of a hydrogen ambient. Once cleaned, the samples are transferred \textit{in vacuo} to an ultra-high vacuum chamber designed specifically for metal evaporation (with a base pressure $< 5e-11$ Torr).

Within this chamber, the nanowire samples are cooled to 85 K (-188$^{\circ}$C) for 2 hours before the onset of tin evaporation. A 15-nm-thick layer of tin is then evaporated from an effusion cell at a growth rate of 7.5 nm/hr, with an evaporation angle close to 60$^{\circ}$ from the sample normal. This shallow evaporation angle facilitates the in-situ formation of the Sn shell.

Following Sn evaporation, while the sample is anticipated to remain at cryogenic temperatures (due to the thermal mass of the molybdenum block), a 3-nm-thick shell of $\mathrm{AlO_x}$ is electron-beam evaporated onto the nanowire sample at normal incidence. Subsequently, the samples are allowed to warm up to room temperature within a vacuum environment. The $\mathrm{AlO_x}$ layer plays a crucial role in preventing the dewetting of Sn into lumpy grains. It also inhibits the oxidation Sn, however, we note that due to the difference in the deposition angles, not all of the Sn shell may be covered by $\mathrm{AlO_x}$.

\subsection{Chemical and Structural Analysis} 

 We use scanning electron microscopy (SEM) to measure the dimensions of the bare InAs nanowires (total length, bottom diameter, top diameter), calculate their density, and evaluate the substrate surface quality (density of craters, visible smoothness) right after InAs nanowire growth. SEM imaging serves also to evaluate the homogeneity and smoothness of the Sn films deposited on both the InAs nanowires and the InAs substrates. The shadow junctions are visualized by SEM on as grown samples and on nanowires transferred on $\mathrm{SiO_2}$ substrates for further device fabrication.
 
Transmission electron microscopy (TEM) is employed to evaluate the structural properties of the Sn shells grown on the InAs nanowires and to evaluate how they form around the nanowires. We use high resolution TEM (HRTEM) to determine the crystalline phase of Sn, while Scanning TEM-High-Angle Annular Dark-Field (STEM-HAADF) and Energy-Dispersive X-ray Spectroscopy (EDS) allow to evaluate the Sn/InAs interface abruptness, determine the size of the junctions and evaluate the coverage of $\mathrm{AlO_x}$ on the Sn shells.


\subsection{Transport Device Fabrication} 

Devices are fabricated in a fashion similar to previous works on wires with Sn shells \cite{pendharkar2019paritypreserving, zhang2022evidence, Bomin2023_QPC}. The device substrates are p-doped Si wafers used as electrostatic back gates. The wafers are covered by 285 nm of thermal silicon oxide, and an additional 10 nm layer of HfOx, which form the back gate dielectric.  The substrates are patterned with gold pads and alignment markers. Nanowires are collected from the growth chip with the help of 300 nm wide micro-manipulator tungsten tips and deterministically placed under an optical microscope onto device chips. Electron-beam lithography is used to pattern the ohmic contacts to nanowires as well as side gate electrodes. The electron beam resist is cured at room temperature under vacuum for at least 8 hours. This is done to avoid nanowire heating and potential interdiffusion of Sn and In. Before evaporating 10/140 nm of Ti/Au to metalize the gates and contacts, an Ar ion milling step is performed to remove the $\mathrm{AlO_x}$ layer. Contacts are completed through a standard lift-off process.


\subsection{Transport Measurements} 
\label{methods:Transport}
To investigate the Sn-induced superconducting transport in InAs nanowires, we prepare field-effect transistor style Josephson-Junction devices as described above and measure them in two-terminal current and voltage bias configurations. We employ a local side gate near the junction and a global back gate to control the charge carrier density within the semiconductor. The measurements are carried out in a dilution refrigerator at approximately 70 mK, utilizing a combination of both direct current and lock-in techniques. The cryostat features a superconducting vector magnet. The raw two-terminal data includes all series resistances in our measurement setup, specifically the $\sim 4.4k\Omega$ resistance from low-pass RC filters in the refrigerator lines, the input and output resistances of the source and readout modules, and $\sim 100\Omega$ drop in the fridge lines, wire bonds and lithographic leads. Full accounting of these resistances is explained in Supplementary Section \ref{SupSec:TwoTermResisance}. 

\section{Results}

\subsection{Criss-crossing InAs nanowire landscapes}

\begin{figure*}
  \includegraphics[page=1, trim=0cm 0cm 0cm 0cm, clip, width=\textwidth]{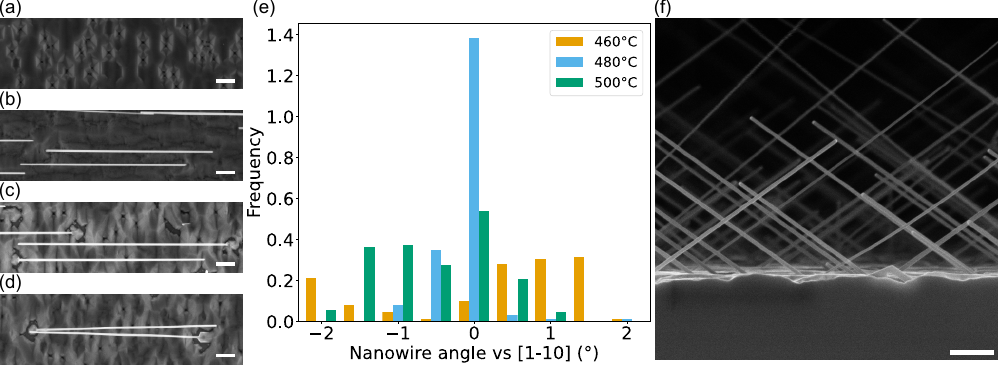}%
  \caption{Inclined InAs nanowires. Top view SEM images of inclined nanowires grown after sample annealing at (a) 733 K (460$^{\circ}$C)
  , (b) 753 K 
  480$^{\circ}$C
  , (c) 773 K
  500$^{\circ}$C 
  and (d) 793 K, respectively. Scale bars 200 nm. (e) Histogram showing the distribution of the nanowires orientation with respect to the in-plane $[1\overline{1}0]$ orientation of the substrate. (f) Side view SEM image of inclined nanowires grown after an annealing temperature of 773 K. Scale bar 500 nm.}
  \label{fig:SEMwires}
  \centering
\end{figure*}

We evaluate the effect of the substrate annealing temperature on the growth of InAs nanowires, as shown in Fig.\ref{fig:SEMwires} (a)-(d). These top view images depict inclined InAs nanowires grown under identical conditions, except for the annealing step where the temperatures are different. We observe that after annealing the sample at 793 K (520$^{\circ}$C), the nanowires do not grow, as shown in Fig.\ref{fig:SEMwires}(a). The surface presents shallow pits, with clearly defined facets, similar to the ones present on homoepitaxial InAs(001) \cite{Ye2013homo}. In the vicinity of these pits,  gold particles  that did not nucleate nanowires are visible (see supplementary Figure \ref{fig_suppl:Annealing}).

For annealing temperatures lower than 793 K (520$^{\circ}$C), all nanowire samples show similar morphology, dimensions and density. The nanowires grown after annealing at 773 K (500$^{\circ}$C) are oriented parallel to each other, with a small percentage ($<1\%$) growing as nanoflakes in the perpendicular direction. When the annealing temperature decreases down to 733 K (460$^{\circ}$C), the nanowire growth axis orientation deviates from the $[1\overline{1}0]$ direction, and nanowires that are close to each other start to touch. The surface of the substrate is a complete yet wavy carpet layer when annealing is performed at 773 K (500$^{\circ}$C). When annealing the substrate at lower temperatures, the carpet film is incomplete and faceted, pits appear with increasing density when the annealing temperature decreases. 

The nanowire misorientation dependence with the annealing temperature is shown in the histogram of Fig.~\ref{fig:SEMwires}(d): for annealing temperatures of 500$^{\circ}$C and below, the occurrence of nanowires with their growth axis parallel to the $[1\overline{1}0]$ direction is high, and is maximal at 753 K (480$^{\circ}$C). To explain the behavior, we looked carefully at the sample grown after annealing at 733 K (460$^{\circ}$C), as it gives the lower parallelism. We find that the nanowires' growth axis deviates from the $[1\overline{1}0]$ direction in several ways : the nanowire nucleates at the edge between a {111}B and a {110} facet of the pedestal, the pedestal is misaligned with respect to the substrate, or the nanowire nucleation is disturbed by spurious 3D nucleation at the base (see zoomed in SEM in supplementary Figure \ref{fig_suppl:Misoriented}). We speculate that the imperfect geometrical definition and misalignment of the pedestal are due to the presence of remaining oxide patches, which reduce the epitaxial relationship between the substrate and the pedestal and disrupt the formation of clear facets on the pedestal. Overall, the annealing temperature affects both nanowire nucleation and alignment, and the annealing temperature range of 40°C, within which wires grow with the right alignment, is acceptable.

A side view SEM image of the InAs nanowires grown using an annealing temperature of 500$^{\circ}$C is shown on Fig.~\ref{fig:SEMwires}(e). The wires grow out of plane with an angle of 35.6$^{\circ}$. The wires grown from 30 nm diameter Au-catalyst measure 6-10 $\mathrm{\mu m}$ long with a diameter ranging between 50 and 60 nm. The dispersion in length is due to the growth being In-flux limited, in which the nanowire length is strongly dependent on the diameter that is tied to the colloid size \cite{mosiiets2024dual}.

In contrast with our results, in earlier studies annealing was performed at higher temperatures, above 773 K \cite{KangReclined2017,KangKshape2017,Krizek2017}. The growth conditions were otherwise similar to ours: the wires were grown in the indium diffusion-limited regime at temperatures ranging between 673 K and 723 K (400$^{\circ}$C to 450$^{\circ}$C). A striking observation is that the nanowire growth time was at least twice as long, despite yielding nanowires of comparable length. Early growth stage was recorded in \cite{KangReclined2017} and in our work (see supplementary Figure \ref{fig_suppl:InitGrowthStage}). In the first case, after 10 minutes of growth, nanowires emerge from craters, whereas in our case, nanowires grow from pedestals after one minute, which disappear within the carpet layer after 10 minutes of growth depending on the prior annealing temperature.

We propose that at high annealing temperatures, indium is incorporated in excess into the gold catalyst from the substrate, which delays the growth. Indeed, the Au droplet is fed without limitation by In coming from the possible decomposition of the substrate as shown by the phase diagram at temperatures above 787 K (514$^{\circ}$C) \cite{Okamoto2004gold,Sung2015gold}.  Due to the excess of In in the droplet, the wetting angle deviates from the ideal value for nanowire nucleation on (111)B surfaces. As long as no (111)B facets are available, the gold droplet  moves randomly on the (001) surface of the substrate, forming craters \cite{Krizek2017,Kang2018}, until (111)B facets form inside a crater. When a droplet reaches such a facet, it stabilizes and evacuates the excess indium via nanowire nucleation and elongation. In our case, i.e at lower annealing temperatures, the In concentration inside the droplet is set by the liquidus\cite{Okamoto2004gold,Sung2015gold}. As long as no (111)B facets are present, nanowire growth occurs along the surface, as it is favored with respect to nucleation at the droplet/substrate interface. A pedestal is formed and once the droplet sits on one of the (111)B facets of the mound, the actual elongation in the (111)B direction takes place.

\subsection{Sn shadow junctions}

\begin{figure}
  \includegraphics[page=1, trim=0cm 0cm 0cm 0cm, clip, width=0.5\textwidth]{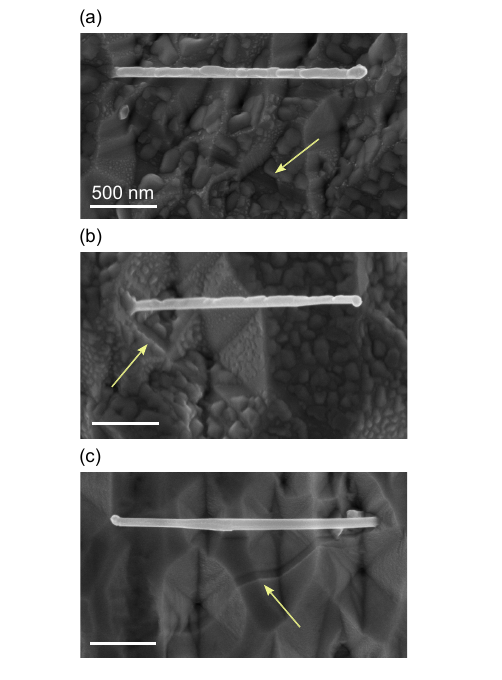}%
  \caption{Sn half-shells grown on inclined InAs nanowires: morphology. Top view SEM images of InAs nanowires covered with a Sn shell deposited at 150 K (a), 120 K (b) and 80 K (c), respectively. The yellow arrows point towards the shadow.}
  \label{fig:SEMshells}
\end{figure}

First, we determine the optimal deposition temperature of Sn to form a smooth shell on the nanowires. Fig.~\ref{fig:SEMshells}(a)-(c) shows top view images of inclined InAs nanowires half-covered by a 10 nm-Sn shell deposited at different cryogenic temperatures. When shell deposition is performed above 150 K (Fig.~\ref{fig:SEMshells}(a) and (b)), it appears granular and inhomogeneous, similar to the work reported by Khan et al. \cite{khan2023}. The Sn grains are visible: they measure hundreds of nanometers in length and are disconnected. These features are present on both the nanowire shells and in the carpet film grown on the substrate. In contrast, when the deposition temperature decreases as low as 80 K, the shell is smooth and homogeneous. The presence of the shell can be confirmed by a sharp shadow below the nanowire. This shadow is present because the Sn evaporation beam is at  60$^{\circ}$ to substrate normal. 

\begin{figure*}
  \includegraphics[page=1, trim=0cm 0cm 0cm 0cm, clip, width=\textwidth]{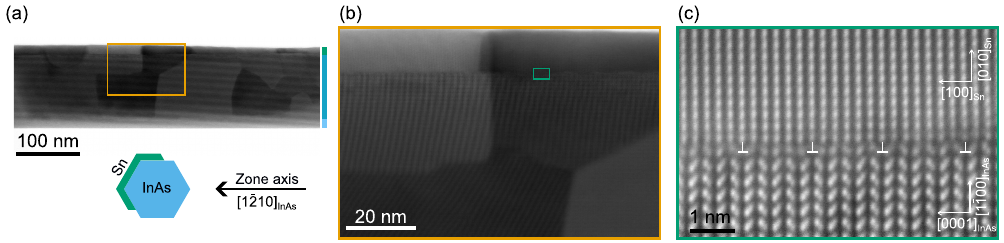}%
  \caption{Sn half-shells on InAs nanowires: structural studies. (a) STEM-HAADF image of a nanowire highlighting the presence of grains in the Sn shell and schematic of the nanowire cross section. (b) Close view of the nanowire-shell interface showing the presence of Moiré patterns. (c) High resolution STEM-HAADF image of the Sn/InAs interface showing the epitaxial relationship between InAs and Sn and the presence of an array of dislocations. All the images are oriented along the $[11\overline{2}0]_{InAs}$ zone axis.}
  \label{fig:TEMshells}
\end{figure*}

We use TEM to analyze the grain size and orientation of the Sn shells that appear to have uniform thickness in SEM. Fig.~\ref{fig:TEMshells}(a) shows a HAADF-STEM image of a single InAs nanowire covered with a 15nm-thick Sn shell deposited at 80 K onto 3 of the InAs facets. The shell is  capped with an amorphous 3nm-thick $\mathrm{AlO_x}$ protective layer\cite{pendharkar2019paritypreserving}. The growth axis is oriented horizontally. The polycrystalline nature of the film is revealed through the alternating bright and dark zones corresponding to different Sn domains. The change in contrast between bright and dark is due to reduced scattering when the grain orientation switches between zone axis of different indices. The dimensions of 44 grains are measured (full set of images available in the data repository). The average lateral dimensions of the Sn grains are $\mathrm{50\pm23 nm}$. These grains have dimensions in line with those reported for epitaxial Al shells on InAs nanowires \cite{Krogstrup2015,KangReclined2017} and polycrystalline Sn shells grown on InSb nanowires \cite{pendharkar2019paritypreserving}. Larger grain sizes may offer advantage for some physics applications such as Majorana Zero Modes, where any disorder is detrimental to the putative topological gap. In other cases, such as for transmon qubits, the effect of the grains is not yet established.

Fig.~\ref{fig:TEMshells}(b) shows a closer view of the Sn/InAs heterostructure, highlighted in Fig.~\ref{fig:TEMshells}(a) by a orange rectangle. Two grains separated by a boundary are visible on top of the nanowire edge. When the shell covers the InAs nanowire, Moir{\'e} patterns emerge due to the overlap between the crystal lattices of InAs and Sn. For three different grains, the fringes of the Moir{\'e} patterns have the same periodicity ($\sim$1.7 nm) and varying angles (see supplementary section \ref{fig_suppl:Moire}). The fringe inclination varies from 54$^{\circ}$ to 86.5$^{\circ}$ with respect to the [0001] direction of wurtzite InAs.

Fig.~\ref{fig:TEMshells}(c) is a high resolution image of the interface between a Sn grain and the InAs nanowire that corresponds to the green rectangle inFig.~\ref{fig:TEMshells}(b). This grain of Sn is an epitaxial relationship with the InAs lattice. By performing a Fast Fourier Transform of the high resolution images, we determine the phase and the orientation of the grain. The tetragonal $\beta$-Sn  grain shown in Fig.~\ref{fig:TEMshells}(c). The $[100]/[010]_{Sn}$  orientations of $\beta$-Sn are aligned with the $[0001]/[10\bar{1}0]_{InAs}$ directions of the InAs nanowire core. A regular array of mismatch dislocations forms at the interface. For every 10 monolayers of InAs, 6-7 monolayers of Sn are accommodated. 

We studied 14 additional Sn grains, whose crystalline planes are visible in high resolution TEM images oriented along either the $[1\bar{2}10]_{InAs}$ or the $[10\bar{1}0]_{InAs}$ zone axis of InAs (See supplementary section \ref{SupSec:GrainTEM}). We find that all the grains  have the $\beta$ crystalline phase. Most of these grains are oriented with $[100]_{Sn}$ parallel to $[0001]_{InAs}$. We find that the $[101]_{Sn}$ direction is also present and is difficult to differentiate from the $[100]_{Sn}$ because the d-spacing of $\{200\}_{Sn}$ and $\{101\}_{Sn}$ planes are very close (292 and 279 pm, respectively). Interestingly, when the InAs nanowire is oriented along the $[10\bar{1}0]$ zone axis, the Sn grains are oriented such that the lattice forms almost a honeycomb with $[100]_{Sn}$ and $[101]_{Sn}$ orientations forming an angle of 62$^{\circ}$.

We evaluate the misorientation of the Sn grains overlapping the InAs nanowire. We start with a hypothesis that on Fig.~\ref{fig:TEMshells}(b), all grains showing a Moir{\'e} have similar orientations. We then simulate the orientation of Moir{\'e} fringes when a Sn grain with its $[100]/[010]_{Sn}$ orientation is rotated relative to the $[0001]/[10\bar{1}0]_{InAs}$ orientation of the InAs nanowire. By rotating the model Sn lattice with respect to the InAs lattice by angles ranging from 1$^{\circ}$ to 7$^{\circ}$, we can reproduce the Moir{\'e} fringes for different grains, which inclination range from 54$^{\circ}$ to 86.5$^{\circ}$.

\begin{figure}
  \includegraphics[page=1, trim=0cm 0cm 0cm 0cm, clip, width=0.45\textwidth]{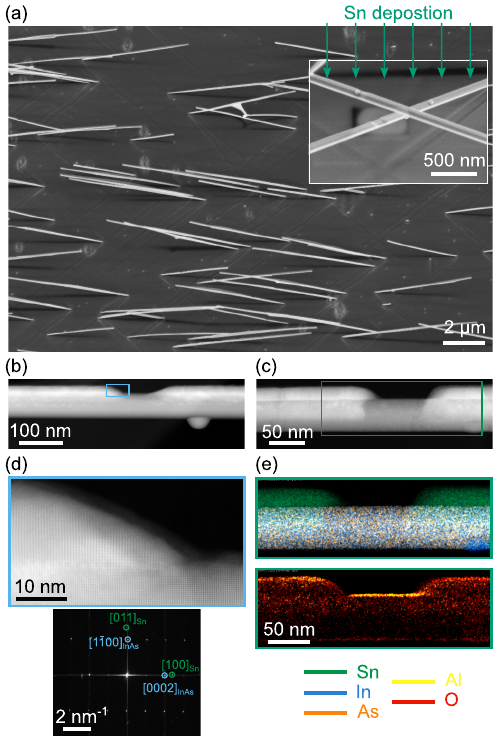}%
  \caption{Sn/InAs/Sn nanowire shadow junctions: structural and chemical analysis. (a) SEM image of inclined InAs nanowires covered with a Sn half shell (10$^{\circ}$ tilt). The inset shows a close view of two nanowires crossing and the presence of the junction on the hind nanowire (20$^{\circ}$ tilt). (b) and (c) HAADF-TEM images of Sn/InAs/Sn nanowire junctions. (d) Close-up view of the junction from (b) and indexed Fast Fourier Transforms of the image. (e) Chemical analysis of the junction. HAADF-EDS color maps indicating Sn (green), In (blue), As (orange), Al (yellow) and O (red).}
  \label{fig:TEMshadow}
\end{figure}

We investigate the crystalline and chemical properties of Sn/InAs/Sn junctions fabricated with the nanowire shadowing technique. In Fig.~\ref{fig:TEMshadow}(a), the wires are inclined along one of the two crystalline orientations. Elongated shadows, formed during the deposition of Sn, are visible on the surface of the substrate, starting from the bottom of each nanowire. The inset shows two nanowires crossing: the back nanowire features a break in the Sn shell due to the shadowing of the Sn beam by the front nanowire, these become superconductor-semiconductor-superconductor junctions at lower temperatures. An optimal nanowire density exists when growing 6 $\mathrm{\mu m}$ long nanowires from colloids, at which junctions form. A density that is too low prevents the formation of criss-crossed nanowires, while a density that is too high results in multiple junctions within a single nanowire. In this particular sample, a density of 0.25 $\mathrm{\mu m^{-2}}$ is ideal for creating single junctions, provided the nanowire distribution across the sample is homogeneous.


Fig.~\ref{fig:TEMshadow}(b) and (c) are HAADF-STEM close-up views of two Sn/InAs/Sn shadow junctions. The length of the break in the Sn shell is around 50 nm and the sloped part of the shell segment is about 30 nm. Fig.~\ref{fig:TEMshadow}(d) is a high resolution image of the edge of the shadow from Fig.~\ref{fig:TEMshadow}(b). Based on the FFT, we find that the in-plane orientation of the grain is similar to the grain in Fig.~\ref{fig:TEMshells}(c), its out-of-plane orientation is $[011]_{Sn}$. The interface is rough, on the order of a few  atomic layers. Fig.~\ref{fig:TEMshadow}(e) is a chemical mapping by EDS of the elements for the junction shown in Fig.~\ref{fig:TEMshadow}(c). The shell is composed of pure Sn and the core of the nanowire is InAs. A droplet of pure In is visible at the bottom right of the nanowire, these droplets are also visible in the inset of panel (a). We speculate that they form during the atomic hydrogen oxyde desorption of the nanowire surface performed prior to Sn deposition. The occurrence of In droplets can be controlled by reducing H exposure. The bottom image in panel (e) shows the presence of the protective $\mathrm{AlO_x}$ cap on the top surface of the Sn/InAs/Sn junction.

The fact that Sn shells grown on our InAs nanowires are comprised only of $\beta$-Sn grains is in contrast with how Sn grows on InSb nanowires \cite{pendharkar2019paritypreserving} and  on InAsSb/InSb nanowire heterostructures \cite{khan2023} whose Sn shells contain also $\alpha$-Sn grains. $\beta$-Sn formation is facilitated by the large lattice mismatch with wurtzite InAs and, according to \cite{khan2023}, by the lack of symmetry between both crystals. 

While $\beta$-Sn grows as a polycrystalline shell, the grains tend to align along a specific direction relative to the nanowire axis when grown on wurtzite InAs. In contrast, in shells grown on Zinc-Blende InSb nanowires, the grains are randomly oriented \cite{pendharkar2019paritypreserving}. The preferential orientation corresponds to the $\mathrm{[100]_{Sn}}$ direction being parallel to the $\mathrm{[0001]_{InAs}}$ direction because the domain formed by 5x$\mathrm{(0002)_{InAs}}$ and 6x$\mathrm{\{200\}_{Sn}}$ planes have a reduced lattice mismatch of 1\%. We also observe some domains formed by 4x$\mathrm{(0002)_{InAs}}$ and 5x$\mathrm{\{101\}_{Sn}}$ planes with a lattice mismatch of 1.5\%. Yet, the network of dislocations is less regular.

Finally, it is interesting to note that the 15-nm-thick shell remains smooth along the nanowire length despite its polycrystalline nature. We attribute this feature to the low thickness of the film, which prevents faceting of the grains~\cite{Krogstrup2015}, combined with the effect of the $\mathrm{AlO_x}$ capping layer. Interestingly, this is the first time such homogeneous layers of Sn have been formed on InAs nanowires, as previous attempts reported that the shells were discontinuous despite elongated grains up to 200nm long and of homogeneous thickness\cite{khan2023}. The fact that our Sn shells are grown after H-cleaning of the InAs nanowire surface might explain the reduced size of our grains by the presence of local roughness. One way to increase their size would be to grow the entire stack $in-situ$. As an alternative, increasing the temperature during growth may contribute to the elongation of the grains while preventing coalescence.

\subsection{Induced Superconductivity}

\begin{figure*}
  \includegraphics[page=1, trim=0cm 0cm 0cm 0cm, clip, width=170mm]{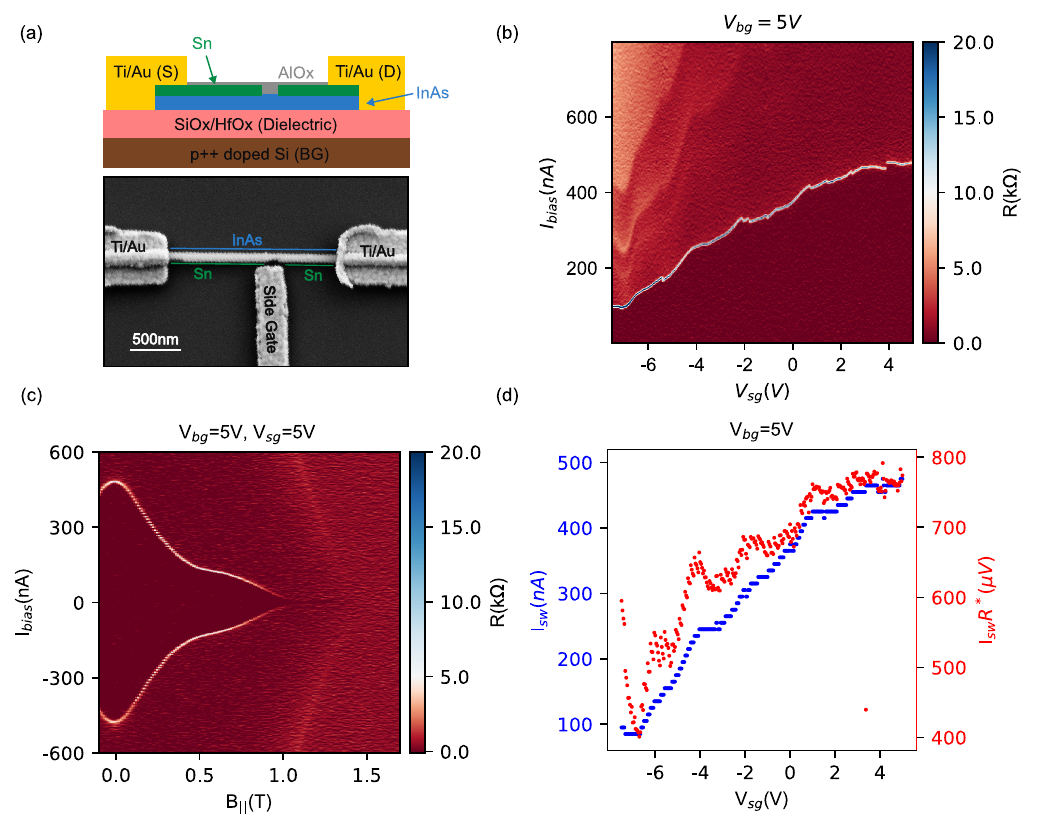}
  \caption{Nanowire Junction Device and Current bias transport measurements (a) Top: vertical cross-section through a device (not to scale), Bottom: SEM image of Device A. Ti/Au leads contact the Sn-InAs nanowire on either ends as Source and Drain terminals while another Ti/Au lead acts as a side gate on top of dielectic covered doped Si substrate serving as a back gate (b) Differential resistance (dV/dI) as a function of current bias and side gate voltage, with the backgate voltage fixed at $V_{bg}=5V$ (C) Differential resistance (dV/dI) as a function of current bias and magnetic field along the nanowire at $V_{bg}=5V, V_{sg}=5V$. (d) Extracted $I_{sw}$ on left axis and $I_{sw}R_n$ product on right axis for different side gate values at $V_{sg}$=-7.5V.}
  \label{fig:Ibias}
\end{figure*}

Having incorporated the Sn-InAs nanowires with shadow junctions into devices for transport measurements, as described in the Methods section \ref{methods:Transport} and shown in Fig.~\ref{fig:Ibias}a, we study the effects of induced superconductivity. In Fig.~\ref{fig:Ibias}b, we track the switching supercurrent $I_{sw}$ as the lowest-current-valued peak in the differential resistance (dV/dI) at different gate voltage combinations. $I_{sw}$ increases with side-gate voltage as the carrier density in the InAs junction grows. With a back gate voltage $V_{bg}=5V$, we record supercurrents of up to 480 nA. The resonances above $I_{sw}$ can be explained by multiple Andreev reflection processes (See supplementary section \ref{SupSec:MARS}). 
In Fig.~\ref{fig:Ibias}c, we look at the evolution of the switching current with a magnetic field along the nanowire $B_{||}$ at $V_{bg} = V_{sg}= 5V$ where $V_{sg}$ is the side gate voltage (See supplementary section \ref{SupSec:OrientB} where we explain the procedure for orienting the magnetic field along nanowire). The bias current is swept from negative to positive values, therefore we observe retrapping currents $I_{ret}$ in the negative current bias direction. A parallel critical magnetic field of approximately 1.3 T is established for this device. The non-monotonic decay of $I_{sw}$ in the multimode regime (high positive gates) with a parallel magnetic field, as observed here, has also been previously reported in \cite{Kun2017_FewModesPaper, Aranya2023_SnInAsSAG}. This behavior could arise from a combination of factors, including but not limited to the interference of multiple transverse propagating modes in the presence of disorder, as well as spin-orbit, orbital, and Zeeman effects in the material. However, a detailed investigation is beyond the scope of the current paper.

 Within -$7.5V \leq V_{sg} \leq 5V$ at $V_{bg} = 5V$, the product $I_{sw}R^*$ varied between 400 - 800 $\mu V$ for device A as shown in right axis of Fig. ~\ref{fig:Ibias}d along with extracted $I_{sw}$ plotted on the left axis. This range is comparable to the superconducting gap of bulk Sn (575 $\mu V$ at 0 K) and approaches the commonly referenced Ambegaokar-Baratoff limit ($903 \mu e V$) in the open regime at higher gate voltage settings. It is also comparable to the findings in other hybrid Sn structures \cite{pendharkar2019paritypreserving, Aranya2023_SnInAsSAG}. We see similar $I_{sw}R^*$ product ranges for other measured devices (see supplementary Table \ref{table:summarySNS}). We approximate the normal state junction resistance $R_n$ as $R^*$ from the linear part of current-voltage traces at high current biases for each gate point individually (see more in supplementary section \ref{SupSec:IswRn}). We sweep the currents high enough to ensure the measured voltage across the device is much greater than $2\Delta$.


$Note:$ The $I_C R_N$ product, computed by combining the critical current $I_C$ with the normal state resistance $R_n$ of the junction, serves as a commonly employed metric for evaluating the weak link quality by comparing it with theroetical predictions by Ambegaokar-Baratoff\cite{Ambegaokar1963_Tunneling}, Kulik-Omel'yanchuk \cite{Kulik1975_JosephsonEffect, Kulik1977_Microbridges} and Likharev \cite{Likharev1979_WeakLinks}. However real junctions can switch to normal state prematurely due to thermal fluctuations \cite{tinkham2004introduction} or supercurrents suppressed due to environment effects \cite{Jarillo-Herrero2006_Supercurrent, 1994_ParitySuppression_Devoret}. The measured switching current can be smaller than the actual critical current and hence $I_{sw} R_N$ products are often reported as a proxy for the $I_C R_N$ \cite{Hertel2021_SAG}.

\begin{figure*}
  \includegraphics[page=1, trim=0cm 0cm 0cm -1cm, clip, width=155mm]{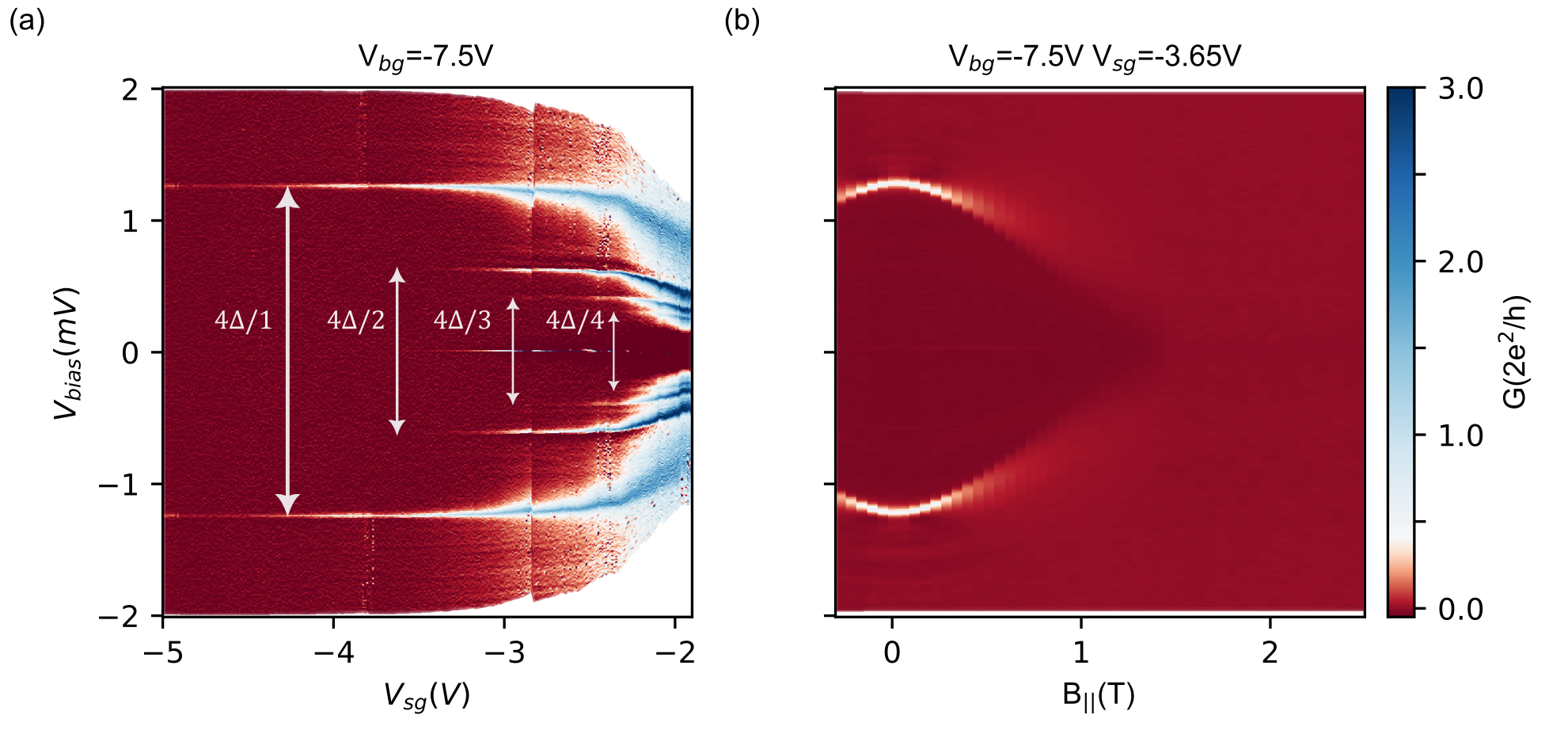}
  \caption{Voltage bias transport measurements. (a) Differential conductance variation with $V_{bias}$ voltage and side gate. $V_{bg}$ = –7.5 V. Finite bias resonances are marked with arrows and are labeled. 
  (b) Magnetic field dependence of the superconducting gap. $V_{bg}$ = -7.5 V, $V_{sg}$ = -3.65 V.}
\label{fig:Vbias}
\end{figure*}

In voltage biased measurements, we aim to study induced superconductivity near the pinch-off regime of the nanowire. In Fig.~\ref{fig:Vbias}(a), the conductance map appears devoid of Coulomb diamond features, which however does not preclude that accidental quantum dots are present. Sub-gap peak-dip resonances that appear at certain voltage bias values are hallmarks of induced superconductivity. The resonance at the highest bias $V_{bias} \approx$ 1.2 mV is assigned to to gap edge tunneling between the two Sn segments, which should appear at  $\pm2\Delta$. The resonances at lower bias values can be attributed to multiple Andreev reflection (MAR) processes. Resonances at zero bias are due to supercurrents that are studied in the current biased measurements.

A superconducting gap $\Delta \approx$ 630 \unit{\uV} is consistent with the series of finite bias resonances, and agrees with the bulk superconducting gap of Sn.  In Fig.~\ref{fig:Vbias}b, with gate voltages set close to the pinch-off regime so that only the 4$\Delta$ resonance is visible, we obtain differential conductance as a function of normalized voltage bias across the device at different parallel magnetic fields. We see that the gap closes with magnetic field but persists to $\mathrm{\sim 1.2~T}$, which is comparable to the critical field for this device obtained in current bias measurements (Fig.~\ref{fig:Ibias}c). Other devices yielded critical fields between $\mathrm{1.1-1.7~T}$ (see supplementary Table \ref{table:summarySNS}).

\section{Conclusions}

We report shadow-defined junctions based on InAs/Sn nanowires. Sn is found to be promising as an alternative to the widely used Al. The inclined nanowire shadowing technique is generic and can be applied to evaluate various superconductor shells. Three key results are discussed:
\begin{itemize}
\item We grow inclined InAs nanowires with two opposite and parallel growth orientations.
\item The Sn shell grows smoothly at 80 K on InAs forming shadow junctions. The shell is in the $\beta$-Sn crystalline phase and has a preferential orientation of $[200]$ relative to the $[0001]$ orientation of InAs.
\item Sn/InAs/Sn Josephson junctions are gate-tunable, with a maximal critical current of 500 nA and parallel critical fields exceeding 1 Tesla.  
\end{itemize}

The large critical currents suggest that Sn-InAs nanowire Josephson junctions are promising candidates for single-nanowire based superconducting microwave circuits such as gate-tunable non-linear inductive elements. For gate-tunable superconducting transmon qubits, the wires result in microsecond scale coherence times.

\begin{acknowledgments}

Nanowire growth was supported by ANR HYBRID (ANR-17-PIRE-0001), ANR IMAGIQUE (ANR-42-PRC-0047), IRP HYNATOQ and the Transatlantic Research Partnership. Sn shell growth was supported by the NSF Quantum Foundry at UCSB funded via the Q-AMASE-i program under award DMR-1906325. TEM characterization was supported by the U.S. Department of Energy through grant DE-SC-0019274. Transport measurements were supported by the U.S. Department of
Energy, Basic Energy Sciences grant DE-SC-0022073.


\end{acknowledgments}

\section*{Duration and Volume of Study}
 The study was spread across 2018-2024. The inclined InAs nanowires for transport studies were grown in 2021 and the nanowires grown for materials studies were grown between 2021 and 2023. Over 300 batches of nanowire samples were grown by MBE. Sn was deposited in January 2023. Two device chips for transport studies were prepared in March 2023. A total of 12 S-NW-S junctions on two chips were measured over two cooldowns between April-September 2023. Approximately 2400 data sets were collected.




\bibliography{apssamp}

\begin{thebibliography}{58}%
\makeatletter
\providecommand \@ifxundefined [1]{%
 \@ifx{#1\undefined}
}%
\providecommand \@ifnum [1]{%
 \ifnum #1\expandafter \@firstoftwo
 \else \expandafter \@secondoftwo
 \fi
}%
\providecommand \@ifx [1]{%
 \ifx #1\expandafter \@firstoftwo
 \else \expandafter \@secondoftwo
 \fi
}%
\providecommand \natexlab [1]{#1}%
\providecommand \enquote  [1]{``#1''}%
\providecommand \bibnamefont  [1]{#1}%
\providecommand \bibfnamefont [1]{#1}%
\providecommand \citenamefont [1]{#1}%
\providecommand \href@noop [0]{\@secondoftwo}%
\providecommand \href [0]{\begingroup \@sanitize@url \@href}%
\providecommand \@href[1]{\@@startlink{#1}\@@href}%
\providecommand \@@href[1]{\endgroup#1\@@endlink}%
\providecommand \@sanitize@url [0]{\catcode `\\12\catcode `\$12\catcode
  `\&12\catcode `\#12\catcode `\^12\catcode `\_12\catcode `\%12\relax}%
\providecommand \@@startlink[1]{}%
\providecommand \@@endlink[0]{}%
\providecommand \url  [0]{\begingroup\@sanitize@url \@url }%
\providecommand \@url [1]{\endgroup\@href {#1}{\urlprefix }}%
\providecommand \urlprefix  [0]{URL }%
\providecommand \Eprint [0]{\href }%
\providecommand \doibase [0]{https://doi.org/}%
\providecommand \selectlanguage [0]{\@gobble}%
\providecommand \bibinfo  [0]{\@secondoftwo}%
\providecommand \bibfield  [0]{\@secondoftwo}%
\providecommand \translation [1]{[#1]}%
\providecommand \BibitemOpen [0]{}%
\providecommand \bibitemStop [0]{}%
\providecommand \bibitemNoStop [0]{.\EOS\space}%
\providecommand \EOS [0]{\spacefactor3000\relax}%
\providecommand \BibitemShut  [1]{\csname bibitem#1\endcsname}%
\let\auto@bib@innerbib\@empty
\bibitem [{\citenamefont {De~Lange}\ \emph {et~al.}(2015)\citenamefont
  {De~Lange}, \citenamefont {Van~Heck}, \citenamefont {Bruno}, \citenamefont
  {Van~Woerkom}, \citenamefont {Geresdi}, \citenamefont {Plissard},
  \citenamefont {Bakkers}, \citenamefont {Akhmerov},\ and\ \citenamefont
  {DiCarlo}}]{LangeDiCarlo2015_Gatemon}%
  \BibitemOpen
  \bibfield  {author} {\bibinfo {author} {\bibfnamefont {G.}~\bibnamefont
  {De~Lange}}, \bibinfo {author} {\bibfnamefont {B.}~\bibnamefont {Van~Heck}},
  \bibinfo {author} {\bibfnamefont {A.}~\bibnamefont {Bruno}}, \bibinfo
  {author} {\bibfnamefont {D.}~\bibnamefont {Van~Woerkom}}, \bibinfo {author}
  {\bibfnamefont {A.}~\bibnamefont {Geresdi}}, \bibinfo {author} {\bibfnamefont
  {S.}~\bibnamefont {Plissard}}, \bibinfo {author} {\bibfnamefont
  {E.}~\bibnamefont {Bakkers}}, \bibinfo {author} {\bibfnamefont
  {A.}~\bibnamefont {Akhmerov}},\ and\ \bibinfo {author} {\bibfnamefont
  {L.}~\bibnamefont {DiCarlo}},\ }\bibfield  {title} {\bibinfo {title}
  {Realization of microwave quantum circuits using hybrid
  superconducting-semiconducting nanowire {Josephson} elements},\ }\href
  {https://doi.org/10.1103/PhysRevLett.115.127002} {\bibfield  {journal}
  {\bibinfo  {journal} {Physical review letters}\ }\textbf {\bibinfo {volume}
  {115}},\ \bibinfo {pages} {127002} (\bibinfo {year} {2015})}\BibitemShut
  {NoStop}%
\bibitem [{\citenamefont {Casparis}\ \emph {et~al.}(2019)\citenamefont
  {Casparis}, \citenamefont {Pearson}, \citenamefont {Kringh{\o}j},
  \citenamefont {Larsen}, \citenamefont {Kuemmeth}, \citenamefont {Nyg{\aa}rd},
  \citenamefont {Krogstrup}, \citenamefont {Petersson},\ and\ \citenamefont
  {Marcus}}]{Casparis2019_coupler}%
  \BibitemOpen
  \bibfield  {author} {\bibinfo {author} {\bibfnamefont {L.}~\bibnamefont
  {Casparis}}, \bibinfo {author} {\bibfnamefont {N.~J.}\ \bibnamefont
  {Pearson}}, \bibinfo {author} {\bibfnamefont {A.}~\bibnamefont
  {Kringh{\o}j}}, \bibinfo {author} {\bibfnamefont {T.~W.}\ \bibnamefont
  {Larsen}}, \bibinfo {author} {\bibfnamefont {F.}~\bibnamefont {Kuemmeth}},
  \bibinfo {author} {\bibfnamefont {J.}~\bibnamefont {Nyg{\aa}rd}}, \bibinfo
  {author} {\bibfnamefont {P.}~\bibnamefont {Krogstrup}}, \bibinfo {author}
  {\bibfnamefont {K.~D.}\ \bibnamefont {Petersson}},\ and\ \bibinfo {author}
  {\bibfnamefont {C.~M.}\ \bibnamefont {Marcus}},\ }\bibfield  {title}
  {\bibinfo {title} {Voltage-controlled superconducting quantum bus},\ }\href
  {https://doi.org/10.1103/PhysRevB.99.085434} {\bibfield  {journal} {\bibinfo
  {journal} {Physical Review B}\ }\textbf {\bibinfo {volume} {99}},\ \bibinfo
  {pages} {085434} (\bibinfo {year} {2019})}\BibitemShut {NoStop}%
\bibitem [{\citenamefont {Hays}\ \emph {et~al.}(2021)\citenamefont {Hays},
  \citenamefont {Fatemi}, \citenamefont {Bouman}, \citenamefont {Cerrillo},
  \citenamefont {Diamond}, \citenamefont {Serniak}, \citenamefont {Connolly},
  \citenamefont {Krogstrup}, \citenamefont {Nyg{\aa}rd}, \citenamefont
  {Levy~Yeyati} \emph {et~al.}}]{HaysDevoret2021_AndreevQubit}%
  \BibitemOpen
  \bibfield  {author} {\bibinfo {author} {\bibfnamefont {M.}~\bibnamefont
  {Hays}}, \bibinfo {author} {\bibfnamefont {V.}~\bibnamefont {Fatemi}},
  \bibinfo {author} {\bibfnamefont {D.}~\bibnamefont {Bouman}}, \bibinfo
  {author} {\bibfnamefont {J.}~\bibnamefont {Cerrillo}}, \bibinfo {author}
  {\bibfnamefont {S.}~\bibnamefont {Diamond}}, \bibinfo {author} {\bibfnamefont
  {K.}~\bibnamefont {Serniak}}, \bibinfo {author} {\bibfnamefont
  {T.}~\bibnamefont {Connolly}}, \bibinfo {author} {\bibfnamefont
  {P.}~\bibnamefont {Krogstrup}}, \bibinfo {author} {\bibfnamefont
  {J.}~\bibnamefont {Nyg{\aa}rd}}, \bibinfo {author} {\bibfnamefont
  {A.}~\bibnamefont {Levy~Yeyati}}, \emph {et~al.},\ }\bibfield  {title}
  {\bibinfo {title} {Coherent manipulation of an andreev spin qubit},\ }\href
  {https://doi.org/10.1126/science.abf0345} {\bibfield  {journal} {\bibinfo
  {journal} {Science}\ }\textbf {\bibinfo {volume} {373}},\ \bibinfo {pages}
  {430} (\bibinfo {year} {2021})}\BibitemShut {NoStop}%
\bibitem [{\citenamefont {Splitthoff}\ \emph {et~al.}(2024)\citenamefont
  {Splitthoff}, \citenamefont {Wesdorp}, \citenamefont {Pita-Vidal},
  \citenamefont {Bargerbos}, \citenamefont {Liu},\ and\ \citenamefont
  {Andersen}}]{splitthoff2024gate}%
  \BibitemOpen
  \bibfield  {author} {\bibinfo {author} {\bibfnamefont {L.~J.}\ \bibnamefont
  {Splitthoff}}, \bibinfo {author} {\bibfnamefont {J.~J.}\ \bibnamefont
  {Wesdorp}}, \bibinfo {author} {\bibfnamefont {M.}~\bibnamefont {Pita-Vidal}},
  \bibinfo {author} {\bibfnamefont {A.}~\bibnamefont {Bargerbos}}, \bibinfo
  {author} {\bibfnamefont {Y.}~\bibnamefont {Liu}},\ and\ \bibinfo {author}
  {\bibfnamefont {C.~K.}\ \bibnamefont {Andersen}},\ }\bibfield  {title}
  {\bibinfo {title} {Gate-tunable kinetic inductance parametric amplifier},\
  }\href {https://doi.org/10.1103/PhysRevApplied.21.014052} {\bibfield
  {journal} {\bibinfo  {journal} {Physical Review Applied}\ }\textbf {\bibinfo
  {volume} {21}},\ \bibinfo {pages} {014052} (\bibinfo {year}
  {2024})}\BibitemShut {NoStop}%
\bibitem [{\citenamefont {Larsen}\ \emph {et~al.}(2015)\citenamefont {Larsen},
  \citenamefont {Petersson}, \citenamefont {Kuemmeth}, \citenamefont
  {Jespersen}, \citenamefont {Krogstrup}, \citenamefont {Nyg{\aa}rd},\ and\
  \citenamefont {Marcus}}]{LarsenMarcus2015_Gatemon}%
  \BibitemOpen
  \bibfield  {author} {\bibinfo {author} {\bibfnamefont {T.~W.}\ \bibnamefont
  {Larsen}}, \bibinfo {author} {\bibfnamefont {K.~D.}\ \bibnamefont
  {Petersson}}, \bibinfo {author} {\bibfnamefont {F.}~\bibnamefont {Kuemmeth}},
  \bibinfo {author} {\bibfnamefont {T.~S.}\ \bibnamefont {Jespersen}}, \bibinfo
  {author} {\bibfnamefont {P.}~\bibnamefont {Krogstrup}}, \bibinfo {author}
  {\bibfnamefont {J.}~\bibnamefont {Nyg{\aa}rd}},\ and\ \bibinfo {author}
  {\bibfnamefont {C.~M.}\ \bibnamefont {Marcus}},\ }\bibfield  {title}
  {\bibinfo {title} {Semiconductor-nanowire-based superconducting qubit},\
  }\href {https://doi.org/10.1103/PhysRevLett.115.127001} {\bibfield  {journal}
  {\bibinfo  {journal} {Physical review letters}\ }\textbf {\bibinfo {volume}
  {115}},\ \bibinfo {pages} {127001} (\bibinfo {year} {2015})}\BibitemShut
  {NoStop}%
\bibitem [{\citenamefont {Hao}\ \emph {et~al.}(2024)\citenamefont {Hao},
  \citenamefont {Shaw}, \citenamefont {Hatefipour}, \citenamefont {Strickland},
  \citenamefont {Elfeky}, \citenamefont {Langone}, \citenamefont {Shabani},\
  and\ \citenamefont {Shankar}}]{hao2024kerr}%
  \BibitemOpen
  \bibfield  {author} {\bibinfo {author} {\bibfnamefont {Z.}~\bibnamefont
  {Hao}}, \bibinfo {author} {\bibfnamefont {T.}~\bibnamefont {Shaw}}, \bibinfo
  {author} {\bibfnamefont {M.}~\bibnamefont {Hatefipour}}, \bibinfo {author}
  {\bibfnamefont {W.}~\bibnamefont {Strickland}}, \bibinfo {author}
  {\bibfnamefont {B.}~\bibnamefont {Elfeky}}, \bibinfo {author} {\bibfnamefont
  {D.}~\bibnamefont {Langone}}, \bibinfo {author} {\bibfnamefont
  {J.}~\bibnamefont {Shabani}},\ and\ \bibinfo {author} {\bibfnamefont
  {S.}~\bibnamefont {Shankar}},\ }\bibfield  {title} {\bibinfo {title} {Kerr
  nonlinearity and parametric amplification with an al-inas
  superconductor-semiconductor {Josephson} junction},\ }\href
  {https://arxiv.org/abs/2402.11085} {\bibfield  {journal} {\bibinfo  {journal}
  {arXiv preprint arXiv:2402.11085}\ } (\bibinfo {year} {2024})}\BibitemShut
  {NoStop}%
\bibitem [{\citenamefont {Phan}\ \emph {et~al.}(2023)\citenamefont {Phan},
  \citenamefont {Falthansl-Scheinecker}, \citenamefont {Mishra}, \citenamefont
  {Strickland}, \citenamefont {Langone}, \citenamefont {Shabani},\ and\
  \citenamefont {Higginbotham}}]{phan2023gate}%
  \BibitemOpen
  \bibfield  {author} {\bibinfo {author} {\bibfnamefont {D.}~\bibnamefont
  {Phan}}, \bibinfo {author} {\bibfnamefont {P.}~\bibnamefont
  {Falthansl-Scheinecker}}, \bibinfo {author} {\bibfnamefont {U.}~\bibnamefont
  {Mishra}}, \bibinfo {author} {\bibfnamefont {W.}~\bibnamefont {Strickland}},
  \bibinfo {author} {\bibfnamefont {D.}~\bibnamefont {Langone}}, \bibinfo
  {author} {\bibfnamefont {J.}~\bibnamefont {Shabani}},\ and\ \bibinfo {author}
  {\bibfnamefont {A.~P.}\ \bibnamefont {Higginbotham}},\ }\bibfield  {title}
  {\bibinfo {title} {Gate-tunable superconductor-semiconductor parametric
  amplifier},\ }\href {https://doi.org/10.1103/PhysRevApplied.19.064032}
  {\bibfield  {journal} {\bibinfo  {journal} {Physical Review Applied}\
  }\textbf {\bibinfo {volume} {19}},\ \bibinfo {pages} {064032} (\bibinfo
  {year} {2023})}\BibitemShut {NoStop}%
\bibitem [{\citenamefont {Sagi}\ \emph {et~al.}(2024)\citenamefont {Sagi},
  \citenamefont {Crippa}, \citenamefont {Valentini}, \citenamefont {Janik},
  \citenamefont {Baghumyan}, \citenamefont {Fabris}, \citenamefont {Kapoor},
  \citenamefont {Hassani}, \citenamefont {Fink}, \citenamefont {Calcaterra},
  \citenamefont {Chrastina}, \citenamefont {Isella},\ and\ \citenamefont
  {Katsaros}}]{sagi_gate_2024}%
  \BibitemOpen
  \bibfield  {author} {\bibinfo {author} {\bibfnamefont {O.}~\bibnamefont
  {Sagi}}, \bibinfo {author} {\bibfnamefont {A.}~\bibnamefont {Crippa}},
  \bibinfo {author} {\bibfnamefont {M.}~\bibnamefont {Valentini}}, \bibinfo
  {author} {\bibfnamefont {M.}~\bibnamefont {Janik}}, \bibinfo {author}
  {\bibfnamefont {L.}~\bibnamefont {Baghumyan}}, \bibinfo {author}
  {\bibfnamefont {G.}~\bibnamefont {Fabris}}, \bibinfo {author} {\bibfnamefont
  {L.}~\bibnamefont {Kapoor}}, \bibinfo {author} {\bibfnamefont
  {F.}~\bibnamefont {Hassani}}, \bibinfo {author} {\bibfnamefont
  {J.}~\bibnamefont {Fink}}, \bibinfo {author} {\bibfnamefont {S.}~\bibnamefont
  {Calcaterra}}, \bibinfo {author} {\bibfnamefont {D.}~\bibnamefont
  {Chrastina}}, \bibinfo {author} {\bibfnamefont {G.}~\bibnamefont {Isella}},\
  and\ \bibinfo {author} {\bibfnamefont {G.}~\bibnamefont {Katsaros}},\
  }\bibfield  {title} {\bibinfo {title} {A gate tunable transmon qubit in
  planar {Ge}},\ }\href {https://doi.org/10.1038/s41467-024-50763-6} {\bibfield
   {journal} {\bibinfo  {journal} {Nature Communications}\ }\textbf {\bibinfo
  {volume} {15}},\ \bibinfo {pages} {6400} (\bibinfo {year}
  {2024})}\BibitemShut {NoStop}%
\bibitem [{\citenamefont {Butseraen}\ \emph {et~al.}(2022)\citenamefont
  {Butseraen}, \citenamefont {Ranadive}, \citenamefont {Aparicio},
  \citenamefont {Rafsanjani~Amin}, \citenamefont {Juyal}, \citenamefont
  {Esposito}, \citenamefont {Watanabe}, \citenamefont {Taniguchi},
  \citenamefont {Roch}, \citenamefont {Lefloch} \emph
  {et~al.}}]{butseraen2022gate}%
  \BibitemOpen
  \bibfield  {author} {\bibinfo {author} {\bibfnamefont {G.}~\bibnamefont
  {Butseraen}}, \bibinfo {author} {\bibfnamefont {A.}~\bibnamefont {Ranadive}},
  \bibinfo {author} {\bibfnamefont {N.}~\bibnamefont {Aparicio}}, \bibinfo
  {author} {\bibfnamefont {K.}~\bibnamefont {Rafsanjani~Amin}}, \bibinfo
  {author} {\bibfnamefont {A.}~\bibnamefont {Juyal}}, \bibinfo {author}
  {\bibfnamefont {M.}~\bibnamefont {Esposito}}, \bibinfo {author}
  {\bibfnamefont {K.}~\bibnamefont {Watanabe}}, \bibinfo {author}
  {\bibfnamefont {T.}~\bibnamefont {Taniguchi}}, \bibinfo {author}
  {\bibfnamefont {N.}~\bibnamefont {Roch}}, \bibinfo {author} {\bibfnamefont
  {F.}~\bibnamefont {Lefloch}}, \emph {et~al.},\ }\bibfield  {title} {\bibinfo
  {title} {A gate-tunable graphene {Josephson} parametric amplifier},\ }\href
  {https://doi.org/10.1038/s41565-022-01235-9} {\bibfield  {journal} {\bibinfo
  {journal} {Nature Nanotechnology}\ }\textbf {\bibinfo {volume} {17}},\
  \bibinfo {pages} {1153} (\bibinfo {year} {2022})}\BibitemShut {NoStop}%
\bibitem [{\citenamefont {Wang}\ \emph {et~al.}(2018)\citenamefont {Wang},
  \citenamefont {Rodan-Legrain}, \citenamefont {Bretheau}, \citenamefont
  {Campbell}, \citenamefont {Kannan}, \citenamefont {Kim}, \citenamefont
  {Kjaergaard}, \citenamefont {Krantz}, \citenamefont {Samach}, \citenamefont
  {Yan}, \citenamefont {Yoder}, \citenamefont {Watanabe}, \citenamefont
  {Taniguchi}, \citenamefont {Orlando}, \citenamefont {Gustavsson},
  \citenamefont {Jarillo-Herrero},\ and\ \citenamefont
  {Oliver}}]{Oliver2018_Graphene}%
  \BibitemOpen
  \bibfield  {author} {\bibinfo {author} {\bibfnamefont {J.~I.-J.}\
  \bibnamefont {Wang}}, \bibinfo {author} {\bibfnamefont {D.}~\bibnamefont
  {Rodan-Legrain}}, \bibinfo {author} {\bibfnamefont {L.}~\bibnamefont
  {Bretheau}}, \bibinfo {author} {\bibfnamefont {D.~L.}\ \bibnamefont
  {Campbell}}, \bibinfo {author} {\bibfnamefont {B.}~\bibnamefont {Kannan}},
  \bibinfo {author} {\bibfnamefont {D.}~\bibnamefont {Kim}}, \bibinfo {author}
  {\bibfnamefont {M.}~\bibnamefont {Kjaergaard}}, \bibinfo {author}
  {\bibfnamefont {P.}~\bibnamefont {Krantz}}, \bibinfo {author} {\bibfnamefont
  {G.~O.}\ \bibnamefont {Samach}}, \bibinfo {author} {\bibfnamefont
  {F.}~\bibnamefont {Yan}}, \bibinfo {author} {\bibfnamefont {J.~L.}\
  \bibnamefont {Yoder}}, \bibinfo {author} {\bibfnamefont {K.}~\bibnamefont
  {Watanabe}}, \bibinfo {author} {\bibfnamefont {T.}~\bibnamefont {Taniguchi}},
  \bibinfo {author} {\bibfnamefont {T.~P.}\ \bibnamefont {Orlando}}, \bibinfo
  {author} {\bibfnamefont {S.}~\bibnamefont {Gustavsson}}, \bibinfo {author}
  {\bibfnamefont {P.}~\bibnamefont {Jarillo-Herrero}},\ and\ \bibinfo {author}
  {\bibfnamefont {W.~D.}\ \bibnamefont {Oliver}},\ }\bibfield  {title}
  {\bibinfo {title} {Coherent control of a hybrid superconducting circuit made
  with graphene-based van der waals heterostructures},\ }\href
  {https://doi.org/10.1038/s41565-018-0329-2} {\bibfield  {journal} {\bibinfo
  {journal} {Nature Nanotechnology}\ } (\bibinfo {year} {2018})}\BibitemShut
  {NoStop}%
\bibitem [{\citenamefont {Antony}\ \emph {et~al.}(2021)\citenamefont {Antony},
  \citenamefont {Gustafsson}, \citenamefont {Ribeill}, \citenamefont {Ware},
  \citenamefont {Rajendran}, \citenamefont {Govia}, \citenamefont {Ohki},
  \citenamefont {Taniguchi}, \citenamefont {Watanabe}, \citenamefont {Hone},\
  and\ \citenamefont {Fong}}]{KCFong2021_vdWQubits}%
  \BibitemOpen
  \bibfield  {author} {\bibinfo {author} {\bibfnamefont {A.}~\bibnamefont
  {Antony}}, \bibinfo {author} {\bibfnamefont {M.~V.}\ \bibnamefont
  {Gustafsson}}, \bibinfo {author} {\bibfnamefont {G.~J.}\ \bibnamefont
  {Ribeill}}, \bibinfo {author} {\bibfnamefont {M.}~\bibnamefont {Ware}},
  \bibinfo {author} {\bibfnamefont {A.}~\bibnamefont {Rajendran}}, \bibinfo
  {author} {\bibfnamefont {L.~C.~G.}\ \bibnamefont {Govia}}, \bibinfo {author}
  {\bibfnamefont {T.~A.}\ \bibnamefont {Ohki}}, \bibinfo {author}
  {\bibfnamefont {T.}~\bibnamefont {Taniguchi}}, \bibinfo {author}
  {\bibfnamefont {K.}~\bibnamefont {Watanabe}}, \bibinfo {author}
  {\bibfnamefont {J.}~\bibnamefont {Hone}},\ and\ \bibinfo {author}
  {\bibfnamefont {K.~C.}\ \bibnamefont {Fong}},\ }\bibfield  {title} {\bibinfo
  {title} {Miniaturizing transmon qubits using van der waals materials},\
  }\href {https://doi.org/10.1021/acs.nanolett.1c04160} {\bibfield  {journal}
  {\bibinfo  {journal} {Nano Letters}\ }\textbf {\bibinfo {volume} {21}},\
  \bibinfo {pages} {10122} (\bibinfo {year} {2021})}\BibitemShut {NoStop}%
\bibitem [{\citenamefont {Sarkar}\ \emph {et~al.}(2022)\citenamefont {Sarkar},
  \citenamefont {Salunkhe}, \citenamefont {Mandal}, \citenamefont {Ghatak},
  \citenamefont {Marchawala}, \citenamefont {Das}, \citenamefont {Watanabe},
  \citenamefont {Taniguchi}, \citenamefont {Vijay},\ and\ \citenamefont
  {Deshmukh}}]{RVijay2022_GrapheneJunction}%
  \BibitemOpen
  \bibfield  {author} {\bibinfo {author} {\bibfnamefont {J.}~\bibnamefont
  {Sarkar}}, \bibinfo {author} {\bibfnamefont {K.~V.}\ \bibnamefont
  {Salunkhe}}, \bibinfo {author} {\bibfnamefont {S.}~\bibnamefont {Mandal}},
  \bibinfo {author} {\bibfnamefont {S.}~\bibnamefont {Ghatak}}, \bibinfo
  {author} {\bibfnamefont {A.~H.}\ \bibnamefont {Marchawala}}, \bibinfo
  {author} {\bibfnamefont {I.}~\bibnamefont {Das}}, \bibinfo {author}
  {\bibfnamefont {K.}~\bibnamefont {Watanabe}}, \bibinfo {author}
  {\bibfnamefont {T.}~\bibnamefont {Taniguchi}}, \bibinfo {author}
  {\bibfnamefont {R.}~\bibnamefont {Vijay}},\ and\ \bibinfo {author}
  {\bibfnamefont {M.~M.}\ \bibnamefont {Deshmukh}},\ }\bibfield  {title}
  {\bibinfo {title} {Quantum noise limited microwave amplification using a
  graphene {Josephson} junction},\ }\href
  {https://doi.org/10.1038/s41565-022-01223-z} {\bibfield  {journal} {\bibinfo
  {journal} {Nature Nanotechnology}\ }\textbf {\bibinfo {volume} {17}},\
  \bibinfo {pages} {1147} (\bibinfo {year} {2022})}\BibitemShut {NoStop}%
\bibitem [{\citenamefont {Carrad}\ \emph {et~al.}(2020)\citenamefont {Carrad},
  \citenamefont {Bjergfelt}, \citenamefont {Kanne}, \citenamefont {Aagesen},
  \citenamefont {Krizek}, \citenamefont {Fiordaliso}, \citenamefont {Johnson},
  \citenamefont {Nyg{\aa}rd},\ and\ \citenamefont
  {Jespersen}}]{CarradJesper2020_AlNbTaV_InAs}%
  \BibitemOpen
  \bibfield  {author} {\bibinfo {author} {\bibfnamefont {D.~J.}\ \bibnamefont
  {Carrad}}, \bibinfo {author} {\bibfnamefont {M.}~\bibnamefont {Bjergfelt}},
  \bibinfo {author} {\bibfnamefont {T.}~\bibnamefont {Kanne}}, \bibinfo
  {author} {\bibfnamefont {M.}~\bibnamefont {Aagesen}}, \bibinfo {author}
  {\bibfnamefont {F.}~\bibnamefont {Krizek}}, \bibinfo {author} {\bibfnamefont
  {E.~M.}\ \bibnamefont {Fiordaliso}}, \bibinfo {author} {\bibfnamefont
  {E.}~\bibnamefont {Johnson}}, \bibinfo {author} {\bibfnamefont
  {J.}~\bibnamefont {Nyg{\aa}rd}},\ and\ \bibinfo {author} {\bibfnamefont
  {T.~S.}\ \bibnamefont {Jespersen}},\ }\bibfield  {title} {\bibinfo {title}
  {Shadow epitaxy for in situ growth of generic semiconductor/superconductor
  hybrids},\ }\href {https://doi.org/10.1002/adma.201908411} {\bibfield
  {journal} {\bibinfo  {journal} {Advanced Materials}\ }\textbf {\bibinfo
  {volume} {32}},\ \bibinfo {pages} {1908411} (\bibinfo {year}
  {2020})}\BibitemShut {NoStop}%
\bibitem [{\citenamefont {G{\"u}nel}\ \emph {et~al.}(2012)\citenamefont
  {G{\"u}nel}, \citenamefont {Batov}, \citenamefont {Hardtdegen}, \citenamefont
  {Sladek}, \citenamefont {Winden}, \citenamefont {Weis}, \citenamefont
  {Panaitov}, \citenamefont {Gr{\"u}tzmacher},\ and\ \citenamefont
  {Sch{\"a}pers}}]{GunelSchapers2012_NbInAs}%
  \BibitemOpen
  \bibfield  {author} {\bibinfo {author} {\bibfnamefont {H.}~\bibnamefont
  {G{\"u}nel}}, \bibinfo {author} {\bibfnamefont {I.}~\bibnamefont {Batov}},
  \bibinfo {author} {\bibfnamefont {H.}~\bibnamefont {Hardtdegen}}, \bibinfo
  {author} {\bibfnamefont {K.}~\bibnamefont {Sladek}}, \bibinfo {author}
  {\bibfnamefont {A.}~\bibnamefont {Winden}}, \bibinfo {author} {\bibfnamefont
  {K.}~\bibnamefont {Weis}}, \bibinfo {author} {\bibfnamefont {G.}~\bibnamefont
  {Panaitov}}, \bibinfo {author} {\bibfnamefont {D.}~\bibnamefont
  {Gr{\"u}tzmacher}},\ and\ \bibinfo {author} {\bibfnamefont {T.}~\bibnamefont
  {Sch{\"a}pers}},\ }\bibfield  {title} {\bibinfo {title} {Supercurrent in
  nb/inas-nanowire/nb {Josephson} junctions},\ }\href
  {https://doi.org/10.1063/1.4745024} {\bibfield  {journal} {\bibinfo
  {journal} {Journal of Applied Physics}\ }\textbf {\bibinfo {volume} {112}},\
  \bibinfo {pages} {034316} (\bibinfo {year} {2012})}\BibitemShut {NoStop}%
\bibitem [{\citenamefont {Perla}\ \emph {et~al.}(2021)\citenamefont {Perla},
  \citenamefont {Fonseka}, \citenamefont {Zellekens}, \citenamefont {Deacon},
  \citenamefont {Han}, \citenamefont {K{\"o}lzer}, \citenamefont
  {M{\"o}rstedt}, \citenamefont {Bennemann}, \citenamefont {Espiari},
  \citenamefont {Ishibashi} \emph {et~al.}}]{PerlaShapers2021_NbInAs}%
  \BibitemOpen
  \bibfield  {author} {\bibinfo {author} {\bibfnamefont {P.}~\bibnamefont
  {Perla}}, \bibinfo {author} {\bibfnamefont {H.~A.}\ \bibnamefont {Fonseka}},
  \bibinfo {author} {\bibfnamefont {P.}~\bibnamefont {Zellekens}}, \bibinfo
  {author} {\bibfnamefont {R.}~\bibnamefont {Deacon}}, \bibinfo {author}
  {\bibfnamefont {Y.}~\bibnamefont {Han}}, \bibinfo {author} {\bibfnamefont
  {J.}~\bibnamefont {K{\"o}lzer}}, \bibinfo {author} {\bibfnamefont
  {T.}~\bibnamefont {M{\"o}rstedt}}, \bibinfo {author} {\bibfnamefont
  {B.}~\bibnamefont {Bennemann}}, \bibinfo {author} {\bibfnamefont
  {A.}~\bibnamefont {Espiari}}, \bibinfo {author} {\bibfnamefont
  {K.}~\bibnamefont {Ishibashi}}, \emph {et~al.},\ }\bibfield  {title}
  {\bibinfo {title} {Fully in situ nb/inas-nanowire {Josephson} junctions by
  selective-area growth and shadow evaporation},\ }\href
  {https://doi.org/10.1039/D0NA00999G} {\bibfield  {journal} {\bibinfo
  {journal} {Nanoscale Advances}\ }\textbf {\bibinfo {volume} {3}},\ \bibinfo
  {pages} {1413} (\bibinfo {year} {2021})}\BibitemShut {NoStop}%
\bibitem [{\citenamefont {Paajaste}\ \emph {et~al.}(2015)\citenamefont
  {Paajaste}, \citenamefont {Amado}, \citenamefont {Roddaro}, \citenamefont
  {Bergeret}, \citenamefont {Ercolani}, \citenamefont {Sorba},\ and\
  \citenamefont {Giazotto}}]{Giazotto2015_PbInAs}%
  \BibitemOpen
  \bibfield  {author} {\bibinfo {author} {\bibfnamefont {J.}~\bibnamefont
  {Paajaste}}, \bibinfo {author} {\bibfnamefont {M.}~\bibnamefont {Amado}},
  \bibinfo {author} {\bibfnamefont {S.}~\bibnamefont {Roddaro}}, \bibinfo
  {author} {\bibfnamefont {F.}~\bibnamefont {Bergeret}}, \bibinfo {author}
  {\bibfnamefont {D.}~\bibnamefont {Ercolani}}, \bibinfo {author}
  {\bibfnamefont {L.}~\bibnamefont {Sorba}},\ and\ \bibinfo {author}
  {\bibfnamefont {F.}~\bibnamefont {Giazotto}},\ }\bibfield  {title} {\bibinfo
  {title} {Pb/inas nanowire {Josephson} junction with high critical current and
  magnetic flux focusing},\ }\href {https://doi.org/10.1021/nl504544s}
  {\bibfield  {journal} {\bibinfo  {journal} {Nano letters}\ }\textbf {\bibinfo
  {volume} {15}},\ \bibinfo {pages} {1803} (\bibinfo {year}
  {2015})}\BibitemShut {NoStop}%
\bibitem [{\citenamefont {Chen}\ \emph
  {et~al.}(2023{\natexlab{a}})\citenamefont {Chen}, \citenamefont {van Driel},
  \citenamefont {Lampadaris}, \citenamefont {Khan}, \citenamefont {Alattallah},
  \citenamefont {Zeng}, \citenamefont {Olsson}, \citenamefont {Dvir},
  \citenamefont {Krogstrup},\ and\ \citenamefont
  {Liu}}]{SabbirKrogstrup2023_PbInSb}%
  \BibitemOpen
  \bibfield  {author} {\bibinfo {author} {\bibfnamefont {Y.}~\bibnamefont
  {Chen}}, \bibinfo {author} {\bibfnamefont {D.}~\bibnamefont {van Driel}},
  \bibinfo {author} {\bibfnamefont {C.}~\bibnamefont {Lampadaris}}, \bibinfo
  {author} {\bibfnamefont {S.~A.}\ \bibnamefont {Khan}}, \bibinfo {author}
  {\bibfnamefont {K.}~\bibnamefont {Alattallah}}, \bibinfo {author}
  {\bibfnamefont {L.}~\bibnamefont {Zeng}}, \bibinfo {author} {\bibfnamefont
  {E.}~\bibnamefont {Olsson}}, \bibinfo {author} {\bibfnamefont
  {T.}~\bibnamefont {Dvir}}, \bibinfo {author} {\bibfnamefont {P.}~\bibnamefont
  {Krogstrup}},\ and\ \bibinfo {author} {\bibfnamefont {Y.}~\bibnamefont
  {Liu}},\ }\bibfield  {title} {\bibinfo {title} {Gate-tunable
  superconductivity in hybrid {InSb--Pb} nanowires},\ }\href
  {https://doi.org/10.1063/5.0155663} {\bibfield  {journal} {\bibinfo
  {journal} {Applied Physics Letters}\ }\textbf {\bibinfo {volume} {123}}
  (\bibinfo {year} {2023}{\natexlab{a}})}\BibitemShut {NoStop}%
\bibitem [{\citenamefont {Spathis}\ \emph {et~al.}(2011)\citenamefont
  {Spathis}, \citenamefont {Biswas}, \citenamefont {Roddaro}, \citenamefont
  {Sorba}, \citenamefont {Giazotto},\ and\ \citenamefont
  {Beltram}}]{Giazotto2011_VInAs}%
  \BibitemOpen
  \bibfield  {author} {\bibinfo {author} {\bibfnamefont {P.}~\bibnamefont
  {Spathis}}, \bibinfo {author} {\bibfnamefont {S.}~\bibnamefont {Biswas}},
  \bibinfo {author} {\bibfnamefont {S.}~\bibnamefont {Roddaro}}, \bibinfo
  {author} {\bibfnamefont {L.}~\bibnamefont {Sorba}}, \bibinfo {author}
  {\bibfnamefont {F.}~\bibnamefont {Giazotto}},\ and\ \bibinfo {author}
  {\bibfnamefont {F.}~\bibnamefont {Beltram}},\ }\bibfield  {title} {\bibinfo
  {title} {Hybrid {InAs} nanowire--vanadium proximity squid},\ }\href
  {https://doi.org/10.1088/0957-4484/22/10/105201} {\bibfield  {journal}
  {\bibinfo  {journal} {Nanotechnology}\ }\textbf {\bibinfo {volume} {22}},\
  \bibinfo {pages} {105201} (\bibinfo {year} {2011})}\BibitemShut {NoStop}%
\bibitem [{\citenamefont {Bjergfelt}\ \emph {et~al.}(2019)\citenamefont
  {Bjergfelt}, \citenamefont {Carrad}, \citenamefont {Kanne}, \citenamefont
  {Aagesen}, \citenamefont {Fiordaliso}, \citenamefont {Johnson}, \citenamefont
  {Shojaei}, \citenamefont {Palmstr{\o}m}, \citenamefont {Krogstrup},
  \citenamefont {Jespersen} \emph {et~al.}}]{Krogstrup2019_VInAs_noJJ}%
  \BibitemOpen
  \bibfield  {author} {\bibinfo {author} {\bibfnamefont {M.}~\bibnamefont
  {Bjergfelt}}, \bibinfo {author} {\bibfnamefont {D.~J.}\ \bibnamefont
  {Carrad}}, \bibinfo {author} {\bibfnamefont {T.}~\bibnamefont {Kanne}},
  \bibinfo {author} {\bibfnamefont {M.}~\bibnamefont {Aagesen}}, \bibinfo
  {author} {\bibfnamefont {E.~M.}\ \bibnamefont {Fiordaliso}}, \bibinfo
  {author} {\bibfnamefont {E.}~\bibnamefont {Johnson}}, \bibinfo {author}
  {\bibfnamefont {B.}~\bibnamefont {Shojaei}}, \bibinfo {author} {\bibfnamefont
  {C.~J.}\ \bibnamefont {Palmstr{\o}m}}, \bibinfo {author} {\bibfnamefont
  {P.}~\bibnamefont {Krogstrup}}, \bibinfo {author} {\bibfnamefont {T.~S.}\
  \bibnamefont {Jespersen}}, \emph {et~al.},\ }\bibfield  {title} {\bibinfo
  {title} {Superconducting vanadium/indium-arsenide hybrid nanowires},\ }\href
  {https://doi.org/10.1088/1361-6528/ab15fc} {\bibfield  {journal} {\bibinfo
  {journal} {Nanotechnology}\ }\textbf {\bibinfo {volume} {30}},\ \bibinfo
  {pages} {294005} (\bibinfo {year} {2019})}\BibitemShut {NoStop}%
\bibitem [{\citenamefont {van Schijndel}\ \emph {et~al.}(2024)\citenamefont
  {van Schijndel}, \citenamefont {McFadden}, \citenamefont {Engel},
  \citenamefont {Dong}, \citenamefont {Yánez-Parreño}, \citenamefont
  {Parthasarathy}, \citenamefont {Simmonds},\ and\ \citenamefont
  {Palmstrøm}}]{vanschijndel2024_Ta}%
  \BibitemOpen
  \bibfield  {author} {\bibinfo {author} {\bibfnamefont {T.~A.~J.}\
  \bibnamefont {van Schijndel}}, \bibinfo {author} {\bibfnamefont {A.~P.}\
  \bibnamefont {McFadden}}, \bibinfo {author} {\bibfnamefont {A.~N.}\
  \bibnamefont {Engel}}, \bibinfo {author} {\bibfnamefont {J.~T.}\ \bibnamefont
  {Dong}}, \bibinfo {author} {\bibfnamefont {W.~J.}\ \bibnamefont
  {Yánez-Parreño}}, \bibinfo {author} {\bibfnamefont {M.}~\bibnamefont
  {Parthasarathy}}, \bibinfo {author} {\bibfnamefont {R.~W.}\ \bibnamefont
  {Simmonds}},\ and\ \bibinfo {author} {\bibfnamefont {C.~J.}\ \bibnamefont
  {Palmstrøm}},\ }\href {https://arxiv.org/abs/2405.12417} {\bibinfo {title}
  {Cryogenic growth of tantalum thin films for low-loss superconducting
  circuits}} (\bibinfo {year} {2024}),\ \Eprint
  {https://arxiv.org/abs/2405.12417} {arXiv:2405.12417 [cond-mat.supr-con]}
  \BibitemShut {NoStop}%
\bibitem [{\citenamefont {Pendharkar}\ \emph {et~al.}(2021)\citenamefont
  {Pendharkar}, \citenamefont {Zhang}, \citenamefont {Wu}, \citenamefont
  {Zarassi}, \citenamefont {Zhang}, \citenamefont {Dempsey}, \citenamefont
  {Lee}, \citenamefont {Harrington}, \citenamefont {Badawy}, \citenamefont
  {Gazibegovic}, \citenamefont {Jung}, \citenamefont {Chen}, \citenamefont
  {Verheijen}, \citenamefont {Hocevar}, \citenamefont {Bakkers}, \citenamefont
  {Palmstrøm},\ and\ \citenamefont {Frolov}}]{pendharkar2019paritypreserving}%
  \BibitemOpen
  \bibfield  {author} {\bibinfo {author} {\bibfnamefont {M.}~\bibnamefont
  {Pendharkar}}, \bibinfo {author} {\bibfnamefont {B.}~\bibnamefont {Zhang}},
  \bibinfo {author} {\bibfnamefont {H.}~\bibnamefont {Wu}}, \bibinfo {author}
  {\bibfnamefont {A.}~\bibnamefont {Zarassi}}, \bibinfo {author} {\bibfnamefont
  {P.}~\bibnamefont {Zhang}}, \bibinfo {author} {\bibfnamefont {C.~P.}\
  \bibnamefont {Dempsey}}, \bibinfo {author} {\bibfnamefont {J.~S.}\
  \bibnamefont {Lee}}, \bibinfo {author} {\bibfnamefont {S.~D.}\ \bibnamefont
  {Harrington}}, \bibinfo {author} {\bibfnamefont {G.}~\bibnamefont {Badawy}},
  \bibinfo {author} {\bibfnamefont {S.}~\bibnamefont {Gazibegovic}}, \bibinfo
  {author} {\bibfnamefont {J.}~\bibnamefont {Jung}}, \bibinfo {author}
  {\bibfnamefont {A.-H.}\ \bibnamefont {Chen}}, \bibinfo {author}
  {\bibfnamefont {M.~A.}\ \bibnamefont {Verheijen}}, \bibinfo {author}
  {\bibfnamefont {M.}~\bibnamefont {Hocevar}}, \bibinfo {author} {\bibfnamefont
  {E.~P. A.~M.}\ \bibnamefont {Bakkers}}, \bibinfo {author} {\bibfnamefont
  {C.~J.}\ \bibnamefont {Palmstrøm}},\ and\ \bibinfo {author} {\bibfnamefont
  {S.~M.}\ \bibnamefont {Frolov}},\ }\bibfield  {title} {\bibinfo {title}
  {Parity-preserving and magnetic field resilient superconductivity in indium
  antimonide nanowires with tin shells},\ }\href
  {https://doi.org/10.1126/science.aba5211} {\bibfield  {journal} {\bibinfo
  {journal} {Science}\ }\textbf {\bibinfo {volume} {372}},\ \bibinfo {pages}
  {508} (\bibinfo {year} {2021})}\BibitemShut {NoStop}%
\bibitem [{\citenamefont {Goswami}\ \emph {et~al.}(2023)\citenamefont
  {Goswami}, \citenamefont {Mudi}, \citenamefont {Dempsey}, \citenamefont
  {Zhang}, \citenamefont {Wu}, \citenamefont {Zhang}, \citenamefont {Mitchell},
  \citenamefont {Lee}, \citenamefont {Frolov},\ and\ \citenamefont
  {Palmstrøm}}]{Aranya2023_SnInAsSAG}%
  \BibitemOpen
  \bibfield  {author} {\bibinfo {author} {\bibfnamefont {A.}~\bibnamefont
  {Goswami}}, \bibinfo {author} {\bibfnamefont {S.~R.}\ \bibnamefont {Mudi}},
  \bibinfo {author} {\bibfnamefont {C.}~\bibnamefont {Dempsey}}, \bibinfo
  {author} {\bibfnamefont {P.}~\bibnamefont {Zhang}}, \bibinfo {author}
  {\bibfnamefont {H.}~\bibnamefont {Wu}}, \bibinfo {author} {\bibfnamefont
  {B.}~\bibnamefont {Zhang}}, \bibinfo {author} {\bibfnamefont {W.~J.}\
  \bibnamefont {Mitchell}}, \bibinfo {author} {\bibfnamefont {J.~S.}\
  \bibnamefont {Lee}}, \bibinfo {author} {\bibfnamefont {S.~M.}\ \bibnamefont
  {Frolov}},\ and\ \bibinfo {author} {\bibfnamefont {C.~J.}\ \bibnamefont
  {Palmstrøm}},\ }\bibfield  {title} {\bibinfo {title} {Sn/{InAs} {Josephson}
  {Junctions} on {Selective} {Area} {Grown} {Nanowires} with in {Situ}
  {Shadowed} {Superconductor} {Evaporation}},\ }\href
  {https://doi.org/10.1021/acs.nanolett.3c01320} {\bibfield  {journal}
  {\bibinfo  {journal} {Nano Letters}\ }\textbf {\bibinfo {volume} {23}},\
  \bibinfo {pages} {7311} (\bibinfo {year} {2023})},\ \bibinfo {note}
  {publisher: American Chemical Society}\BibitemShut {NoStop}%
\bibitem [{\citenamefont {Zhang}\ \emph
  {et~al.}(2022{\natexlab{a}})\citenamefont {Zhang}, \citenamefont {Zarassi},
  \citenamefont {Pendharkar}, \citenamefont {Lee}, \citenamefont {Jarjat},
  \citenamefont {Van~de Sande}, \citenamefont {Zhang}, \citenamefont {Mudi},
  \citenamefont {Wu}, \citenamefont {Tan} \emph {et~al.}}]{Po2022_SnSmashJJ}%
  \BibitemOpen
  \bibfield  {author} {\bibinfo {author} {\bibfnamefont {P.}~\bibnamefont
  {Zhang}}, \bibinfo {author} {\bibfnamefont {A.}~\bibnamefont {Zarassi}},
  \bibinfo {author} {\bibfnamefont {M.}~\bibnamefont {Pendharkar}}, \bibinfo
  {author} {\bibfnamefont {J.}~\bibnamefont {Lee}}, \bibinfo {author}
  {\bibfnamefont {L.}~\bibnamefont {Jarjat}}, \bibinfo {author} {\bibfnamefont
  {V.}~\bibnamefont {Van~de Sande}}, \bibinfo {author} {\bibfnamefont
  {B.}~\bibnamefont {Zhang}}, \bibinfo {author} {\bibfnamefont
  {S.}~\bibnamefont {Mudi}}, \bibinfo {author} {\bibfnamefont {H.}~\bibnamefont
  {Wu}}, \bibinfo {author} {\bibfnamefont {S.}~\bibnamefont {Tan}}, \emph
  {et~al.},\ }\bibfield  {title} {\bibinfo {title} {Planar {Josephson}
  junctions templated by nanowire shadowing},\ }\href
  {https://arxiv.org/abs/2211.04130} {\bibfield  {journal} {\bibinfo  {journal}
  {arXiv preprint arXiv:2211.04130}\ } (\bibinfo {year}
  {2022}{\natexlab{a}})}\BibitemShut {NoStop}%
\bibitem [{\citenamefont {Zhang}\ \emph {et~al.}(2024)\citenamefont {Zhang},
  \citenamefont {Zarassi}, \citenamefont {Jarjat}, \citenamefont {Van~de
  Sande}, \citenamefont {Pendharkar}, \citenamefont {Lee}, \citenamefont
  {Dempsey}, \citenamefont {McFadden}, \citenamefont {Harrington},
  \citenamefont {Dong} \emph {et~al.}}]{Po2024_SnSecondOrderJJ}%
  \BibitemOpen
  \bibfield  {author} {\bibinfo {author} {\bibfnamefont {P.}~\bibnamefont
  {Zhang}}, \bibinfo {author} {\bibfnamefont {A.}~\bibnamefont {Zarassi}},
  \bibinfo {author} {\bibfnamefont {L.}~\bibnamefont {Jarjat}}, \bibinfo
  {author} {\bibfnamefont {V.}~\bibnamefont {Van~de Sande}}, \bibinfo {author}
  {\bibfnamefont {M.}~\bibnamefont {Pendharkar}}, \bibinfo {author}
  {\bibfnamefont {J.}~\bibnamefont {Lee}}, \bibinfo {author} {\bibfnamefont
  {C.~P.}\ \bibnamefont {Dempsey}}, \bibinfo {author} {\bibfnamefont
  {A.}~\bibnamefont {McFadden}}, \bibinfo {author} {\bibfnamefont {S.~D.}\
  \bibnamefont {Harrington}}, \bibinfo {author} {\bibfnamefont {J.~T.}\
  \bibnamefont {Dong}}, \emph {et~al.},\ }\bibfield  {title} {\bibinfo {title}
  {Large second-order {Josephson} effect in planar superconductor-semiconductor
  junctions},\ }\href {https://doi.org/10.21468/SciPostPhys.16.1.030}
  {\bibfield  {journal} {\bibinfo  {journal} {SciPost Physics}\ }\textbf
  {\bibinfo {volume} {16}},\ \bibinfo {pages} {030} (\bibinfo {year}
  {2024})}\BibitemShut {NoStop}%
\bibitem [{\citenamefont {Shen}\ \emph {et~al.}(2018)\citenamefont {Shen},
  \citenamefont {Heedt}, \citenamefont {Borsoi}, \citenamefont {van Heck},
  \citenamefont {Gazibegovic}, \citenamefont {Op~het Veld}, \citenamefont
  {Car}, \citenamefont {Logan}, \citenamefont {Pendharkar}, \citenamefont
  {Ramakers}, \citenamefont {Wang}, \citenamefont {Xu}, \citenamefont {Bouman},
  \citenamefont {Geresdi}, \citenamefont {Palmstrøm}, \citenamefont
  {Bakkers},\ and\ \citenamefont {Kouwenhoven}}]{shen_parity_2018}%
  \BibitemOpen
  \bibfield  {author} {\bibinfo {author} {\bibfnamefont {J.}~\bibnamefont
  {Shen}}, \bibinfo {author} {\bibfnamefont {S.}~\bibnamefont {Heedt}},
  \bibinfo {author} {\bibfnamefont {F.}~\bibnamefont {Borsoi}}, \bibinfo
  {author} {\bibfnamefont {B.}~\bibnamefont {van Heck}}, \bibinfo {author}
  {\bibfnamefont {S.}~\bibnamefont {Gazibegovic}}, \bibinfo {author}
  {\bibfnamefont {R.~L.~M.}\ \bibnamefont {Op~het Veld}}, \bibinfo {author}
  {\bibfnamefont {D.}~\bibnamefont {Car}}, \bibinfo {author} {\bibfnamefont
  {J.~A.}\ \bibnamefont {Logan}}, \bibinfo {author} {\bibfnamefont
  {M.}~\bibnamefont {Pendharkar}}, \bibinfo {author} {\bibfnamefont {S.~J.~J.}\
  \bibnamefont {Ramakers}}, \bibinfo {author} {\bibfnamefont {G.}~\bibnamefont
  {Wang}}, \bibinfo {author} {\bibfnamefont {D.}~\bibnamefont {Xu}}, \bibinfo
  {author} {\bibfnamefont {D.}~\bibnamefont {Bouman}}, \bibinfo {author}
  {\bibfnamefont {A.}~\bibnamefont {Geresdi}}, \bibinfo {author} {\bibfnamefont
  {C.~J.}\ \bibnamefont {Palmstrøm}}, \bibinfo {author} {\bibfnamefont {E.~P.
  A.~M.}\ \bibnamefont {Bakkers}},\ and\ \bibinfo {author} {\bibfnamefont
  {L.~P.}\ \bibnamefont {Kouwenhoven}},\ }\bibfield  {title} {\bibinfo {title}
  {Parity transitions in the superconducting ground state of hybrid
  {InSb}–{Al} {Coulomb} islands},\ }\href
  {https://doi.org/10.1038/s41467-018-07279-7} {\bibfield  {journal} {\bibinfo
  {journal} {Nature Communications}\ }\textbf {\bibinfo {volume} {9}},\
  \bibinfo {pages} {4801} (\bibinfo {year} {2018})}\BibitemShut {NoStop}%
\bibitem [{\citenamefont {Abay}\ \emph {et~al.}(2014)\citenamefont {Abay},
  \citenamefont {Persson}, \citenamefont {Nilsson}, \citenamefont {Wu},
  \citenamefont {Xu}, \citenamefont {Fogelstr{\"o}m}, \citenamefont
  {Shumeiko},\ and\ \citenamefont {Delsing}}]{Delsing2014_InAs}%
  \BibitemOpen
  \bibfield  {author} {\bibinfo {author} {\bibfnamefont {S.}~\bibnamefont
  {Abay}}, \bibinfo {author} {\bibfnamefont {D.}~\bibnamefont {Persson}},
  \bibinfo {author} {\bibfnamefont {H.}~\bibnamefont {Nilsson}}, \bibinfo
  {author} {\bibfnamefont {F.}~\bibnamefont {Wu}}, \bibinfo {author}
  {\bibfnamefont {H.}~\bibnamefont {Xu}}, \bibinfo {author} {\bibfnamefont
  {M.}~\bibnamefont {Fogelstr{\"o}m}}, \bibinfo {author} {\bibfnamefont
  {V.}~\bibnamefont {Shumeiko}},\ and\ \bibinfo {author} {\bibfnamefont
  {P.}~\bibnamefont {Delsing}},\ }\bibfield  {title} {\bibinfo {title} {Charge
  transport in {InAs} nanowire {Josephson} junctions},\ }\href
  {https://doi.org/10.1103/PhysRevB.89.214508} {\bibfield  {journal} {\bibinfo
  {journal} {Physical Review B}\ }\textbf {\bibinfo {volume} {89}},\ \bibinfo
  {pages} {214508} (\bibinfo {year} {2014})}\BibitemShut {NoStop}%
\bibitem [{\citenamefont {Wei}\ \emph {et~al.}(2016)\citenamefont {Wei},
  \citenamefont {Katmis}, \citenamefont {Chang},\ and\ \citenamefont
  {Moodera}}]{Moodera2016_V}%
  \BibitemOpen
  \bibfield  {author} {\bibinfo {author} {\bibfnamefont {P.}~\bibnamefont
  {Wei}}, \bibinfo {author} {\bibfnamefont {F.}~\bibnamefont {Katmis}},
  \bibinfo {author} {\bibfnamefont {C.-Z.}\ \bibnamefont {Chang}},\ and\
  \bibinfo {author} {\bibfnamefont {J.~S.}\ \bibnamefont {Moodera}},\
  }\bibfield  {title} {\bibinfo {title} {Induced superconductivity and
  engineered {Josephson} tunneling devices in epitaxial (111)-oriented
  gold/vanadium heterostructures},\ }\href
  {https://doi.org/10.1021/acs.nanolett.6b00114} {\bibfield  {journal}
  {\bibinfo  {journal} {Nano letters}\ }\textbf {\bibinfo {volume} {16}},\
  \bibinfo {pages} {2714} (\bibinfo {year} {2016})}\BibitemShut {NoStop}%
\bibitem [{\citenamefont {G{\"u}sken}\ \emph {et~al.}(2017)\citenamefont
  {G{\"u}sken}, \citenamefont {Rieger}, \citenamefont {Zellekens},
  \citenamefont {Bennemann}, \citenamefont {Neumann}, \citenamefont {Lepsa},
  \citenamefont {Sch{\"a}pers},\ and\ \citenamefont
  {Gr{\"u}tzmacher}}]{Schapers2017_LowTemp}%
  \BibitemOpen
  \bibfield  {author} {\bibinfo {author} {\bibfnamefont {N.~A.}\ \bibnamefont
  {G{\"u}sken}}, \bibinfo {author} {\bibfnamefont {T.}~\bibnamefont {Rieger}},
  \bibinfo {author} {\bibfnamefont {P.}~\bibnamefont {Zellekens}}, \bibinfo
  {author} {\bibfnamefont {B.}~\bibnamefont {Bennemann}}, \bibinfo {author}
  {\bibfnamefont {E.}~\bibnamefont {Neumann}}, \bibinfo {author} {\bibfnamefont
  {M.~I.}\ \bibnamefont {Lepsa}}, \bibinfo {author} {\bibfnamefont
  {T.}~\bibnamefont {Sch{\"a}pers}},\ and\ \bibinfo {author} {\bibfnamefont
  {D.}~\bibnamefont {Gr{\"u}tzmacher}},\ }\bibfield  {title} {\bibinfo {title}
  {Mbe growth of {Al/InA}s and {Nb/InAs} superconducting hybrid nanowire
  structures},\ }\href {https://doi.org/10.1039/C7NR03982D} {\bibfield
  {journal} {\bibinfo  {journal} {Nanoscale}\ }\textbf {\bibinfo {volume}
  {9}},\ \bibinfo {pages} {16735} (\bibinfo {year} {2017})}\BibitemShut
  {NoStop}%
\bibitem [{\citenamefont {Kanne}\ \emph {et~al.}(2021)\citenamefont {Kanne},
  \citenamefont {Marnauza}, \citenamefont {Olsteins}, \citenamefont {Carrad},
  \citenamefont {Sestoft}, \citenamefont {De~Bruijckere}, \citenamefont {Zeng},
  \citenamefont {Johnson}, \citenamefont {Olsson}, \citenamefont
  {Grove-Rasmussen} \emph {et~al.}}]{Jesper2021_PbLowTemp}%
  \BibitemOpen
  \bibfield  {author} {\bibinfo {author} {\bibfnamefont {T.}~\bibnamefont
  {Kanne}}, \bibinfo {author} {\bibfnamefont {M.}~\bibnamefont {Marnauza}},
  \bibinfo {author} {\bibfnamefont {D.}~\bibnamefont {Olsteins}}, \bibinfo
  {author} {\bibfnamefont {D.~J.}\ \bibnamefont {Carrad}}, \bibinfo {author}
  {\bibfnamefont {J.~E.}\ \bibnamefont {Sestoft}}, \bibinfo {author}
  {\bibfnamefont {J.}~\bibnamefont {De~Bruijckere}}, \bibinfo {author}
  {\bibfnamefont {L.}~\bibnamefont {Zeng}}, \bibinfo {author} {\bibfnamefont
  {E.}~\bibnamefont {Johnson}}, \bibinfo {author} {\bibfnamefont
  {E.}~\bibnamefont {Olsson}}, \bibinfo {author} {\bibfnamefont
  {K.}~\bibnamefont {Grove-Rasmussen}}, \emph {et~al.},\ }\bibfield  {title}
  {\bibinfo {title} {Epitaxial {Pb} on {InAs} nanowires for quantum devices},\
  }\href {https://doi.org/10.1038/s41565-021-00892-2} {\bibfield  {journal}
  {\bibinfo  {journal} {Nature Nanotechnology}\ }\textbf {\bibinfo {volume}
  {16}},\ \bibinfo {pages} {776} (\bibinfo {year} {2021})}\BibitemShut
  {NoStop}%
\bibitem [{\citenamefont {Khan}\ \emph {et~al.}(2020)\citenamefont {Khan},
  \citenamefont {Lampadaris}, \citenamefont {Cui}, \citenamefont {Stampfer},
  \citenamefont {Liu}, \citenamefont {Pauka}, \citenamefont {Cachaza},
  \citenamefont {Fiordaliso}, \citenamefont {Kang}, \citenamefont {Korneychuk}
  \emph {et~al.}}]{Sabbir2020_SnTransparentGatable}%
  \BibitemOpen
  \bibfield  {author} {\bibinfo {author} {\bibfnamefont {S.~A.}\ \bibnamefont
  {Khan}}, \bibinfo {author} {\bibfnamefont {C.}~\bibnamefont {Lampadaris}},
  \bibinfo {author} {\bibfnamefont {A.}~\bibnamefont {Cui}}, \bibinfo {author}
  {\bibfnamefont {L.}~\bibnamefont {Stampfer}}, \bibinfo {author}
  {\bibfnamefont {Y.}~\bibnamefont {Liu}}, \bibinfo {author} {\bibfnamefont
  {S.}~\bibnamefont {Pauka}}, \bibinfo {author} {\bibfnamefont {M.~E.}\
  \bibnamefont {Cachaza}}, \bibinfo {author} {\bibfnamefont {E.~M.}\
  \bibnamefont {Fiordaliso}}, \bibinfo {author} {\bibfnamefont {J.-H.}\
  \bibnamefont {Kang}}, \bibinfo {author} {\bibfnamefont {S.}~\bibnamefont
  {Korneychuk}}, \emph {et~al.},\ }\bibfield  {title} {\bibinfo {title} {Highly
  transparent gatable superconducting shadow junctions},\ }\href
  {https://doi.org/10.1021/acsnano.0c02979} {\bibfield  {journal} {\bibinfo
  {journal} {ACS Nano}\ }\textbf {\bibinfo {volume} {14}},\ \bibinfo {pages}
  {14605} (\bibinfo {year} {2020})}\BibitemShut {NoStop}%
\bibitem [{\citenamefont {Khan}\ \emph {et~al.}(2023)\citenamefont {Khan},
  \citenamefont {Martí-Sánchez}, \citenamefont {Olsteins}, \citenamefont
  {Lampadaris}, \citenamefont {Carrad}, \citenamefont {Liu}, \citenamefont
  {Qui{\~n}ones}, \citenamefont {Chiara~Spadaro}, \citenamefont
  {Sand~Jespersen}, \citenamefont {Krogstrup},\ and\ \citenamefont
  {Arbiol}}]{khan2023}%
  \BibitemOpen
  \bibfield  {author} {\bibinfo {author} {\bibfnamefont {S.~A.}\ \bibnamefont
  {Khan}}, \bibinfo {author} {\bibfnamefont {S.}~\bibnamefont
  {Martí-Sánchez}}, \bibinfo {author} {\bibfnamefont {D.}~\bibnamefont
  {Olsteins}}, \bibinfo {author} {\bibfnamefont {C.}~\bibnamefont
  {Lampadaris}}, \bibinfo {author} {\bibfnamefont {D.~J.}\ \bibnamefont
  {Carrad}}, \bibinfo {author} {\bibfnamefont {Y.}~\bibnamefont {Liu}},
  \bibinfo {author} {\bibfnamefont {J.}~\bibnamefont {Qui{\~n}ones}}, \bibinfo
  {author} {\bibfnamefont {M.}~\bibnamefont {Chiara~Spadaro}}, \bibinfo
  {author} {\bibfnamefont {T.}~\bibnamefont {Sand~Jespersen}}, \bibinfo
  {author} {\bibfnamefont {P.}~\bibnamefont {Krogstrup}},\ and\ \bibinfo
  {author} {\bibfnamefont {J.}~\bibnamefont {Arbiol}},\ }\bibfield  {title}
  {\bibinfo {title} {Epitaxially driven phase selectivity of sn in hybrid
  quantum nanowires},\ }\href {https://doi.org/10.1021/acsnano.3c02733}
  {\bibfield  {journal} {\bibinfo  {journal} {ACS Nano}\ }\textbf {\bibinfo
  {volume} {17}},\ \bibinfo {pages} {11794} (\bibinfo {year} {2023})},\
  \bibinfo {note} {pMID: 37317984}\BibitemShut {NoStop}%
\bibitem [{\citenamefont {Chen}\ \emph
  {et~al.}(2023{\natexlab{b}})\citenamefont {Chen}, \citenamefont {Dempsey},
  \citenamefont {Pendharkar}, \citenamefont {Sharma}, \citenamefont {Zhang},
  \citenamefont {Tan}, \citenamefont {Bellon}, \citenamefont {Frolov},
  \citenamefont {Palmstr{\o}m}, \citenamefont {Bellet-Amalric} \emph
  {et~al.}}]{Jane2023_SnCappingLayer}%
  \BibitemOpen
  \bibfield  {author} {\bibinfo {author} {\bibfnamefont {A.-H.}\ \bibnamefont
  {Chen}}, \bibinfo {author} {\bibfnamefont {C.}~\bibnamefont {Dempsey}},
  \bibinfo {author} {\bibfnamefont {M.}~\bibnamefont {Pendharkar}}, \bibinfo
  {author} {\bibfnamefont {A.}~\bibnamefont {Sharma}}, \bibinfo {author}
  {\bibfnamefont {B.}~\bibnamefont {Zhang}}, \bibinfo {author} {\bibfnamefont
  {S.}~\bibnamefont {Tan}}, \bibinfo {author} {\bibfnamefont {L.}~\bibnamefont
  {Bellon}}, \bibinfo {author} {\bibfnamefont {S.~M.}\ \bibnamefont {Frolov}},
  \bibinfo {author} {\bibfnamefont {C.~J.}\ \bibnamefont {Palmstr{\o}m}},
  \bibinfo {author} {\bibfnamefont {E.}~\bibnamefont {Bellet-Amalric}}, \emph
  {et~al.},\ }\bibfield  {title} {\bibinfo {title} {Role of a capping layer on
  the crystalline structure of {Sn} thin films grown at cryogenic temperatures
  on {InSb} substrates},\ }\href {https://doi.org/10.1088/1361-6528/ad079e}
  {\bibfield  {journal} {\bibinfo  {journal} {Nanotechnology}\ }\textbf
  {\bibinfo {volume} {35}},\ \bibinfo {pages} {075702} (\bibinfo {year}
  {2023}{\natexlab{b}})}\BibitemShut {NoStop}%
\bibitem [{\citenamefont {Krogstrup}\ \emph {et~al.}(2015)\citenamefont
  {Krogstrup}, \citenamefont {Ziino}, \citenamefont {Chang}, \citenamefont
  {Albrecht}, \citenamefont {Madsen}, \citenamefont {Johnson}, \citenamefont
  {Nyg{\aa}rd}, \citenamefont {Marcus},\ and\ \citenamefont
  {Jespersen}}]{Krogstrup2015}%
  \BibitemOpen
  \bibfield  {author} {\bibinfo {author} {\bibfnamefont {P.}~\bibnamefont
  {Krogstrup}}, \bibinfo {author} {\bibfnamefont {N.~L.}\ \bibnamefont
  {Ziino}}, \bibinfo {author} {\bibfnamefont {W.}~\bibnamefont {Chang}},
  \bibinfo {author} {\bibfnamefont {S.~M.}\ \bibnamefont {Albrecht}}, \bibinfo
  {author} {\bibfnamefont {M.~H.}\ \bibnamefont {Madsen}}, \bibinfo {author}
  {\bibfnamefont {E.}~\bibnamefont {Johnson}}, \bibinfo {author} {\bibfnamefont
  {J.}~\bibnamefont {Nyg{\aa}rd}}, \bibinfo {author} {\bibfnamefont {C.~M.}\
  \bibnamefont {Marcus}},\ and\ \bibinfo {author} {\bibfnamefont {T.~S.}\
  \bibnamefont {Jespersen}},\ }\bibfield  {title} {\bibinfo {title} {{Epitaxy
  of semiconductor-superconductor nanowires}},\ }\href
  {https://doi.org/10.1038/nmat4176} {\bibfield  {journal} {\bibinfo  {journal}
  {Nature Materials}\ }\textbf {\bibinfo {volume} {14}},\ \bibinfo {pages}
  {400} (\bibinfo {year} {2015})}\BibitemShut {NoStop}%
\bibitem [{\citenamefont {Zeng}\ \emph {et~al.}(2020)\citenamefont {Zeng},
  \citenamefont {Yu}, \citenamefont {Fonseka}, \citenamefont {Boras},
  \citenamefont {Jurczak}, \citenamefont {Wang}, \citenamefont {Sanchez},\ and\
  \citenamefont {Liu}}]{Zeng2020}%
  \BibitemOpen
  \bibfield  {author} {\bibinfo {author} {\bibfnamefont {H.}~\bibnamefont
  {Zeng}}, \bibinfo {author} {\bibfnamefont {X.}~\bibnamefont {Yu}}, \bibinfo
  {author} {\bibfnamefont {H.~A.}\ \bibnamefont {Fonseka}}, \bibinfo {author}
  {\bibfnamefont {G.}~\bibnamefont {Boras}}, \bibinfo {author} {\bibfnamefont
  {P.}~\bibnamefont {Jurczak}}, \bibinfo {author} {\bibfnamefont
  {T.}~\bibnamefont {Wang}}, \bibinfo {author} {\bibfnamefont {A.~M.}\
  \bibnamefont {Sanchez}},\ and\ \bibinfo {author} {\bibfnamefont
  {H.}~\bibnamefont {Liu}},\ }\bibfield  {title} {\bibinfo {title} {{Preferred
  growth direction of III-V nanowires on differently oriented Si substrates}},\
  }\bibfield  {journal} {\bibinfo  {journal} {Nanotechnology}\ }\textbf
  {\bibinfo {volume} {31}},\ \href {https://doi.org/10.1088/1361-6528/abafd7}
  {10.1088/1361-6528/abafd7} (\bibinfo {year} {2020})\BibitemShut {NoStop}%
\bibitem [{\citenamefont {Chen}\ \emph {et~al.}(2016)\citenamefont {Chen},
  \citenamefont {Kapadia}, \citenamefont {Harker}, \citenamefont {Desai},
  \citenamefont {Seuk~Kang}, \citenamefont {Chuang}, \citenamefont {Tosun},
  \citenamefont {Sutter-Fella}, \citenamefont {Tsang}, \citenamefont {Zeng}
  \emph {et~al.}}]{chen2016direct}%
  \BibitemOpen
  \bibfield  {author} {\bibinfo {author} {\bibfnamefont {K.}~\bibnamefont
  {Chen}}, \bibinfo {author} {\bibfnamefont {R.}~\bibnamefont {Kapadia}},
  \bibinfo {author} {\bibfnamefont {A.}~\bibnamefont {Harker}}, \bibinfo
  {author} {\bibfnamefont {S.}~\bibnamefont {Desai}}, \bibinfo {author}
  {\bibfnamefont {J.}~\bibnamefont {Seuk~Kang}}, \bibinfo {author}
  {\bibfnamefont {S.}~\bibnamefont {Chuang}}, \bibinfo {author} {\bibfnamefont
  {M.}~\bibnamefont {Tosun}}, \bibinfo {author} {\bibfnamefont {C.~M.}\
  \bibnamefont {Sutter-Fella}}, \bibinfo {author} {\bibfnamefont
  {M.}~\bibnamefont {Tsang}}, \bibinfo {author} {\bibfnamefont
  {Y.}~\bibnamefont {Zeng}}, \emph {et~al.},\ }\bibfield  {title} {\bibinfo
  {title} {Direct growth of single-crystalline {III--V} semiconductors on
  amorphous substrates},\ }\href@noop {} {\bibfield  {journal} {\bibinfo
  {journal} {Nature communications}\ }\textbf {\bibinfo {volume} {7}},\
  \bibinfo {pages} {10502} (\bibinfo {year} {2016})}\BibitemShut {NoStop}%
\bibitem [{\citenamefont {Dubrovskii}\ \emph {et~al.}(2016)\citenamefont
  {Dubrovskii}, \citenamefont {Berdnikov}, \citenamefont {Schmidtbauer},
  \citenamefont {Borg}, \citenamefont {Storm}, \citenamefont {Deppert},\ and\
  \citenamefont {Johansson}}]{dubrovskii2016length}%
  \BibitemOpen
  \bibfield  {author} {\bibinfo {author} {\bibfnamefont {V.~G.}\ \bibnamefont
  {Dubrovskii}}, \bibinfo {author} {\bibfnamefont {Y.}~\bibnamefont
  {Berdnikov}}, \bibinfo {author} {\bibfnamefont {J.}~\bibnamefont
  {Schmidtbauer}}, \bibinfo {author} {\bibfnamefont {M.}~\bibnamefont {Borg}},
  \bibinfo {author} {\bibfnamefont {K.}~\bibnamefont {Storm}}, \bibinfo
  {author} {\bibfnamefont {K.}~\bibnamefont {Deppert}},\ and\ \bibinfo {author}
  {\bibfnamefont {J.}~\bibnamefont {Johansson}},\ }\bibfield  {title} {\bibinfo
  {title} {Length distributions of nanowires growing by surface diffusion},\
  }\href {https://doi.org/10.1021/acs.cgd.5b01832} {\bibfield  {journal}
  {\bibinfo  {journal} {Crystal Growth \& Design}\ }\textbf {\bibinfo {volume}
  {16}},\ \bibinfo {pages} {2167} (\bibinfo {year} {2016})}\BibitemShut
  {NoStop}%
\bibitem [{\citenamefont {Fr{\"o}berg}\ \emph {et~al.}(2007)\citenamefont
  {Fr{\"o}berg}, \citenamefont {Seifert},\ and\ \citenamefont
  {Johansson}}]{froberg2007diameter}%
  \BibitemOpen
  \bibfield  {author} {\bibinfo {author} {\bibfnamefont {L.}~\bibnamefont
  {Fr{\"o}berg}}, \bibinfo {author} {\bibfnamefont {W.}~\bibnamefont
  {Seifert}},\ and\ \bibinfo {author} {\bibfnamefont {J.}~\bibnamefont
  {Johansson}},\ }\bibfield  {title} {\bibinfo {title} {Diameter-dependent
  growth rate of {InAs} nanowires},\ }\href
  {https://doi.org/10.1103/PhysRevB.76.153401} {\bibfield  {journal} {\bibinfo
  {journal} {Physical Review B}\ }\textbf {\bibinfo {volume} {76}},\ \bibinfo
  {pages} {153401} (\bibinfo {year} {2007})}\BibitemShut {NoStop}%
\bibitem [{\citenamefont {Mosiiets}\ \emph {et~al.}(2024)\citenamefont
  {Mosiiets}, \citenamefont {Genuist}, \citenamefont {Cibert}, \citenamefont
  {Bellet-Amalric},\ and\ \citenamefont {Hocevar}}]{mosiiets2024dual}%
  \BibitemOpen
  \bibfield  {author} {\bibinfo {author} {\bibfnamefont {D.}~\bibnamefont
  {Mosiiets}}, \bibinfo {author} {\bibfnamefont {Y.}~\bibnamefont {Genuist}},
  \bibinfo {author} {\bibfnamefont {J.}~\bibnamefont {Cibert}}, \bibinfo
  {author} {\bibfnamefont {E.}~\bibnamefont {Bellet-Amalric}},\ and\ \bibinfo
  {author} {\bibfnamefont {M.}~\bibnamefont {Hocevar}},\ }\bibfield  {title}
  {\bibinfo {title} {Dual-adatom diffusion-limited growth model for compound
  nanowires: Application to {InAs} nanowires},\ }\bibfield  {journal} {\bibinfo
   {journal} {Crystal Growth \& Design}\ }\href
  {https://doi.org/10.1021/acs.cgd.4c00186} {10.1021/acs.cgd.4c00186} (\bibinfo
  {year} {2024})\BibitemShut {NoStop}%
\bibitem [{\citenamefont {Balakrishnan}\ \emph {et~al.}(2006)\citenamefont
  {Balakrishnan}, \citenamefont {Tatebayashi}, \citenamefont {Khoshakhlagh},
  \citenamefont {Huang}, \citenamefont {Jallipalli}, \citenamefont {Dawson},\
  and\ \citenamefont {Huffaker}}]{Balakrishnan2006}%
  \BibitemOpen
  \bibfield  {author} {\bibinfo {author} {\bibfnamefont {G.}~\bibnamefont
  {Balakrishnan}}, \bibinfo {author} {\bibfnamefont {J.}~\bibnamefont
  {Tatebayashi}}, \bibinfo {author} {\bibfnamefont {A.}~\bibnamefont
  {Khoshakhlagh}}, \bibinfo {author} {\bibfnamefont {S.~H.}\ \bibnamefont
  {Huang}}, \bibinfo {author} {\bibfnamefont {A.}~\bibnamefont {Jallipalli}},
  \bibinfo {author} {\bibfnamefont {L.~R.}\ \bibnamefont {Dawson}},\ and\
  \bibinfo {author} {\bibfnamefont {D.~L.}\ \bibnamefont {Huffaker}},\
  }\bibfield  {title} {\bibinfo {title} {{III/V ratio based selectivity between
  strained Stranski-Krastanov and strain-free GaSb quantum dots on GaAs}},\
  }\href {https://doi.org/10.1063/1.2362999} {\bibfield  {journal} {\bibinfo
  {journal} {Applied Physics Letters}\ }\textbf {\bibinfo {volume} {89}},\
  \bibinfo {pages} {161104} (\bibinfo {year} {2006})},\ \Eprint
  {https://arxiv.org/abs/https://pubs.aip.org/aip/apl/article-pdf/doi/10.1063/1.2362999/14661499/161104\_1\_online.pdf}
  {https://pubs.aip.org/aip/apl/article-pdf/doi/10.1063/1.2362999/14661499/161104\_1\_online.pdf}
  \BibitemShut {NoStop}%
\bibitem [{\citenamefont {Zhang}\ \emph
  {et~al.}(2022{\natexlab{b}})\citenamefont {Zhang}, \citenamefont {Li},
  \citenamefont {Aguilar}, \citenamefont {Zhang}, \citenamefont {Pendharkar},
  \citenamefont {Dempsey}, \citenamefont {Lee}, \citenamefont {Harrington},
  \citenamefont {Tan}, \citenamefont {Meyer} \emph
  {et~al.}}]{zhang2022evidence}%
  \BibitemOpen
  \bibfield  {author} {\bibinfo {author} {\bibfnamefont {B.}~\bibnamefont
  {Zhang}}, \bibinfo {author} {\bibfnamefont {Z.}~\bibnamefont {Li}}, \bibinfo
  {author} {\bibfnamefont {V.}~\bibnamefont {Aguilar}}, \bibinfo {author}
  {\bibfnamefont {P.}~\bibnamefont {Zhang}}, \bibinfo {author} {\bibfnamefont
  {M.}~\bibnamefont {Pendharkar}}, \bibinfo {author} {\bibfnamefont
  {C.}~\bibnamefont {Dempsey}}, \bibinfo {author} {\bibfnamefont
  {J.}~\bibnamefont {Lee}}, \bibinfo {author} {\bibfnamefont {S.}~\bibnamefont
  {Harrington}}, \bibinfo {author} {\bibfnamefont {S.}~\bibnamefont {Tan}},
  \bibinfo {author} {\bibfnamefont {J.}~\bibnamefont {Meyer}}, \emph {et~al.},\
  }\bibfield  {title} {\bibinfo {title} {Evidence of {$\phi 0$-Josephson}
  junction from skewed diffraction patterns in {Sn-InSb} nanowires},\
  }\bibfield  {journal} {\bibinfo  {journal} {arXiv preprint arXiv:2212.00199}\
  }\href {https://doi.org/10.48550/arXiv.2212.00199}
  {10.48550/arXiv.2212.00199} (\bibinfo {year}
  {2022}{\natexlab{b}})\BibitemShut {NoStop}%
\bibitem [{\citenamefont {Zhang}\ \emph {et~al.}(2023)\citenamefont {Zhang},
  \citenamefont {Li}, \citenamefont {Wu}, \citenamefont {Pendharkar},
  \citenamefont {Dempsey}, \citenamefont {Lee}, \citenamefont {Harrington},
  \citenamefont {Palmstrom},\ and\ \citenamefont {Frolov}}]{Bomin2023_QPC}%
  \BibitemOpen
  \bibfield  {author} {\bibinfo {author} {\bibfnamefont {B.}~\bibnamefont
  {Zhang}}, \bibinfo {author} {\bibfnamefont {Z.}~\bibnamefont {Li}}, \bibinfo
  {author} {\bibfnamefont {H.}~\bibnamefont {Wu}}, \bibinfo {author}
  {\bibfnamefont {M.}~\bibnamefont {Pendharkar}}, \bibinfo {author}
  {\bibfnamefont {C.}~\bibnamefont {Dempsey}}, \bibinfo {author} {\bibfnamefont
  {J.}~\bibnamefont {Lee}}, \bibinfo {author} {\bibfnamefont {S.}~\bibnamefont
  {Harrington}}, \bibinfo {author} {\bibfnamefont {C.}~\bibnamefont
  {Palmstrom}},\ and\ \bibinfo {author} {\bibfnamefont {S.}~\bibnamefont
  {Frolov}},\ }\bibfield  {title} {\bibinfo {title} {Supercurrent through a
  single transverse mode in nanowire josephson junctions},\ }\bibfield
  {journal} {\bibinfo  {journal} {arXiv preprint arXiv:2306.00146}\ }\href
  {https://doi.org/10.48550/arXiv.2306.00146} {10.48550/arXiv.2306.00146}
  (\bibinfo {year} {2023})\BibitemShut {NoStop}%
\bibitem [{\citenamefont {Ye}\ \emph {et~al.}(2013)\citenamefont {Ye},
  \citenamefont {Li}, \citenamefont {Hinkey}, \citenamefont {Yang},
  \citenamefont {Mishima}, \citenamefont {Keay}, \citenamefont {Santos},\ and\
  \citenamefont {Johnson}}]{Ye2013homo}%
  \BibitemOpen
  \bibfield  {author} {\bibinfo {author} {\bibfnamefont {H.}~\bibnamefont
  {Ye}}, \bibinfo {author} {\bibfnamefont {L.}~\bibnamefont {Li}}, \bibinfo
  {author} {\bibfnamefont {R.~T.}\ \bibnamefont {Hinkey}}, \bibinfo {author}
  {\bibfnamefont {R.~Q.}\ \bibnamefont {Yang}}, \bibinfo {author}
  {\bibfnamefont {T.~D.}\ \bibnamefont {Mishima}}, \bibinfo {author}
  {\bibfnamefont {J.~C.}\ \bibnamefont {Keay}}, \bibinfo {author}
  {\bibfnamefont {M.~B.}\ \bibnamefont {Santos}},\ and\ \bibinfo {author}
  {\bibfnamefont {M.~B.}\ \bibnamefont {Johnson}},\ }\bibfield  {title}
  {\bibinfo {title} {{MBE growth optimization of {InAs} (001) homoepitaxy}},\
  }\href {https://doi.org/10.1116/1.4804397} {\bibfield  {journal} {\bibinfo
  {journal} {Journal of Vacuum Science {\&} Technology B, Nanotechnology and
  Microelectronics: Materials, Processing, Measurement, and Phenomena}\
  }\textbf {\bibinfo {volume} {31}},\ \bibinfo {pages} {03C135} (\bibinfo
  {year} {2013})}\BibitemShut {NoStop}%
\bibitem [{\citenamefont {Kang}\ \emph
  {et~al.}(2017{\natexlab{a}})\citenamefont {Kang}, \citenamefont {Grivnin},
  \citenamefont {Bor}, \citenamefont {Reiner}, \citenamefont {Avraham},
  \citenamefont {Ronen}, \citenamefont {Cohen}, \citenamefont {Kacman},
  \citenamefont {Shtrikman},\ and\ \citenamefont
  {Beidenkopf}}]{KangReclined2017}%
  \BibitemOpen
  \bibfield  {author} {\bibinfo {author} {\bibfnamefont {J.-H.}\ \bibnamefont
  {Kang}}, \bibinfo {author} {\bibfnamefont {A.}~\bibnamefont {Grivnin}},
  \bibinfo {author} {\bibfnamefont {E.}~\bibnamefont {Bor}}, \bibinfo {author}
  {\bibfnamefont {J.}~\bibnamefont {Reiner}}, \bibinfo {author} {\bibfnamefont
  {N.}~\bibnamefont {Avraham}}, \bibinfo {author} {\bibfnamefont
  {Y.}~\bibnamefont {Ronen}}, \bibinfo {author} {\bibfnamefont
  {Y.}~\bibnamefont {Cohen}}, \bibinfo {author} {\bibfnamefont
  {P.}~\bibnamefont {Kacman}}, \bibinfo {author} {\bibfnamefont
  {H.}~\bibnamefont {Shtrikman}},\ and\ \bibinfo {author} {\bibfnamefont
  {H.}~\bibnamefont {Beidenkopf}},\ }\bibfield  {title} {\bibinfo {title}
  {Robust epitaxial al coating of reclined {InAs} nanowires},\ }\href
  {https://doi.org/10.1021/acs.nanolett.7b03444} {\bibfield  {journal}
  {\bibinfo  {journal} {Nano Letters}\ }\textbf {\bibinfo {volume} {17}},\
  \bibinfo {pages} {7520} (\bibinfo {year} {2017}{\natexlab{a}})},\ \bibinfo
  {note} {pMID: 29115842}\BibitemShut {NoStop}%
\bibitem [{\citenamefont {Kang}\ \emph
  {et~al.}(2017{\natexlab{b}})\citenamefont {Kang}, \citenamefont {Galicka},
  \citenamefont {Kacman},\ and\ \citenamefont {Shtrikman}}]{KangKshape2017}%
  \BibitemOpen
  \bibfield  {author} {\bibinfo {author} {\bibfnamefont {J.-H.}\ \bibnamefont
  {Kang}}, \bibinfo {author} {\bibfnamefont {M.}~\bibnamefont {Galicka}},
  \bibinfo {author} {\bibfnamefont {P.}~\bibnamefont {Kacman}},\ and\ \bibinfo
  {author} {\bibfnamefont {H.}~\bibnamefont {Shtrikman}},\ }\bibfield  {title}
  {\bibinfo {title} {Wurtzite/zinc-blende ‘k’-shape {InAs} nanowires with
  embedded two-dimensional wurtzite plates},\ }\href
  {https://doi.org/10.1021/acs.nanolett.6b04598} {\bibfield  {journal}
  {\bibinfo  {journal} {Nano Letters}\ }\textbf {\bibinfo {volume} {17}},\
  \bibinfo {pages} {531} (\bibinfo {year} {2017}{\natexlab{b}})},\ \bibinfo
  {note} {pMID: 28002676}\BibitemShut {NoStop}%
\bibitem [{\citenamefont {Krizek}\ \emph {et~al.}(2017)\citenamefont {Krizek},
  \citenamefont {Kanne}, \citenamefont {Razmadze}, \citenamefont {Johnson},
  \citenamefont {Nygård}, \citenamefont {Marcus},\ and\ \citenamefont
  {Krogstrup}}]{Krizek2017}%
  \BibitemOpen
  \bibfield  {author} {\bibinfo {author} {\bibfnamefont {F.}~\bibnamefont
  {Krizek}}, \bibinfo {author} {\bibfnamefont {T.}~\bibnamefont {Kanne}},
  \bibinfo {author} {\bibfnamefont {D.}~\bibnamefont {Razmadze}}, \bibinfo
  {author} {\bibfnamefont {E.}~\bibnamefont {Johnson}}, \bibinfo {author}
  {\bibfnamefont {J.}~\bibnamefont {Nygård}}, \bibinfo {author} {\bibfnamefont
  {C.~M.}\ \bibnamefont {Marcus}},\ and\ \bibinfo {author} {\bibfnamefont
  {P.}~\bibnamefont {Krogstrup}},\ }\bibfield  {title} {\bibinfo {title}
  {Growth of {InAs} wurtzite nanocrosses from hexagonal and cubic basis},\
  }\href {https://doi.org/10.1021/acs.nanolett.7b02604} {\bibfield  {journal}
  {\bibinfo  {journal} {Nano Letters}\ }\textbf {\bibinfo {volume} {17}},\
  \bibinfo {pages} {6090} (\bibinfo {year} {2017})},\ \bibinfo {note} {pMID:
  28895746}\BibitemShut {NoStop}%
\bibitem [{\citenamefont {Okamoto}(2004)}]{Okamoto2004gold}%
  \BibitemOpen
  \bibfield  {author} {\bibinfo {author} {\bibfnamefont {H.}~\bibnamefont
  {Okamoto}},\ }\bibfield  {title} {\bibinfo {title} {Au-in (gold-indium)},\
  }\href@noop {} {\bibfield  {journal} {\bibinfo  {journal} {Journal of Phase
  Equilibria and Diffusion}\ }\textbf {\bibinfo {volume} {25}},\ \bibinfo
  {pages} {197} (\bibinfo {year} {2004})}\BibitemShut {NoStop}%
\bibitem [{\citenamefont {Sung}\ \emph {et~al.}(2015)\citenamefont {Sung},
  \citenamefont {Wang},\ and\ \citenamefont {Kim}}]{Sung2015gold}%
  \BibitemOpen
  \bibfield  {author} {\bibinfo {author} {\bibfnamefont {H.~K.}\ \bibnamefont
  {Sung}}, \bibinfo {author} {\bibfnamefont {C.}~\bibnamefont {Wang}},\ and\
  \bibinfo {author} {\bibfnamefont {N.~Y.}\ \bibnamefont {Kim}},\ }\bibfield
  {title} {\bibinfo {title} {{Reliability study of Au-in solid-liquid
  interdiffusion bonding for GaN-based vertical LED packaging}},\ }\href
  {https://doi.org/10.1088/0960-1317/25/12/127002} {\bibfield  {journal}
  {\bibinfo  {journal} {Journal of Micromechanics and Microengineering}\
  }\textbf {\bibinfo {volume} {25}},\ \bibinfo {pages} {127002} (\bibinfo
  {year} {2015})}\BibitemShut {NoStop}%
\bibitem [{\citenamefont {Kang}\ \emph {et~al.}(2018)\citenamefont {Kang},
  \citenamefont {Krizek}, \citenamefont {Zaluska-Kotur}, \citenamefont
  {Krogstrup}, \citenamefont {Kacman}, \citenamefont {Beidenkopf},\ and\
  \citenamefont {Shtrikman}}]{Kang2018}%
  \BibitemOpen
  \bibfield  {author} {\bibinfo {author} {\bibfnamefont {J.-H.}\ \bibnamefont
  {Kang}}, \bibinfo {author} {\bibfnamefont {F.}~\bibnamefont {Krizek}},
  \bibinfo {author} {\bibfnamefont {M.}~\bibnamefont {Zaluska-Kotur}}, \bibinfo
  {author} {\bibfnamefont {P.}~\bibnamefont {Krogstrup}}, \bibinfo {author}
  {\bibfnamefont {P.}~\bibnamefont {Kacman}}, \bibinfo {author} {\bibfnamefont
  {H.}~\bibnamefont {Beidenkopf}},\ and\ \bibinfo {author} {\bibfnamefont
  {H.}~\bibnamefont {Shtrikman}},\ }\bibfield  {title} {\bibinfo {title}
  {Au-assisted substrate-faceting for inclined nanowire growth},\ }\href
  {https://doi.org/10.1021/acs.nanolett.8b00853} {\bibfield  {journal}
  {\bibinfo  {journal} {Nano Letters}\ }\textbf {\bibinfo {volume} {18}},\
  \bibinfo {pages} {4115} (\bibinfo {year} {2018})},\ \bibinfo {note} {pMID:
  29879360}\BibitemShut {NoStop}%
\bibitem [{\citenamefont {Zuo}\ \emph {et~al.}(2017)\citenamefont {Zuo},
  \citenamefont {Mourik}, \citenamefont {Szombati}, \citenamefont {Nijholt},
  \citenamefont {van Woerkom}, \citenamefont {Geresdi}, \citenamefont {Chen},
  \citenamefont {Ostroukh}, \citenamefont {Akhmerov}, \citenamefont {Plissard},
  \citenamefont {Car}, \citenamefont {Bakkers}, \citenamefont {Pikulin},
  \citenamefont {Kouwenhoven},\ and\ \citenamefont
  {Frolov}}]{Kun2017_FewModesPaper}%
  \BibitemOpen
  \bibfield  {author} {\bibinfo {author} {\bibfnamefont {K.}~\bibnamefont
  {Zuo}}, \bibinfo {author} {\bibfnamefont {V.}~\bibnamefont {Mourik}},
  \bibinfo {author} {\bibfnamefont {D.~B.}\ \bibnamefont {Szombati}}, \bibinfo
  {author} {\bibfnamefont {B.}~\bibnamefont {Nijholt}}, \bibinfo {author}
  {\bibfnamefont {D.~J.}\ \bibnamefont {van Woerkom}}, \bibinfo {author}
  {\bibfnamefont {A.}~\bibnamefont {Geresdi}}, \bibinfo {author} {\bibfnamefont
  {J.}~\bibnamefont {Chen}}, \bibinfo {author} {\bibfnamefont {V.~P.}\
  \bibnamefont {Ostroukh}}, \bibinfo {author} {\bibfnamefont {A.~R.}\
  \bibnamefont {Akhmerov}}, \bibinfo {author} {\bibfnamefont {S.~R.}\
  \bibnamefont {Plissard}}, \bibinfo {author} {\bibfnamefont {D.}~\bibnamefont
  {Car}}, \bibinfo {author} {\bibfnamefont {E.~P. A.~M.}\ \bibnamefont
  {Bakkers}}, \bibinfo {author} {\bibfnamefont {D.~I.}\ \bibnamefont
  {Pikulin}}, \bibinfo {author} {\bibfnamefont {L.~P.}\ \bibnamefont
  {Kouwenhoven}},\ and\ \bibinfo {author} {\bibfnamefont {S.~M.}\ \bibnamefont
  {Frolov}},\ }\bibfield  {title} {\bibinfo {title} {Supercurrent interference
  in few-mode nanowire {Josephson} junctions},\ }\href
  {https://doi.org/10.1103/PhysRevLett.119.187704} {\bibfield  {journal}
  {\bibinfo  {journal} {Physical Review Letters}\ }\textbf {\bibinfo {volume}
  {119}},\ \bibinfo {pages} {187704} (\bibinfo {year} {2017})}\BibitemShut
  {NoStop}%
\bibitem [{\citenamefont {Ambegaokar}\ and\ \citenamefont
  {Baratoff}(1963)}]{Ambegaokar1963_Tunneling}%
  \BibitemOpen
  \bibfield  {author} {\bibinfo {author} {\bibfnamefont {V.}~\bibnamefont
  {Ambegaokar}}\ and\ \bibinfo {author} {\bibfnamefont {A.}~\bibnamefont
  {Baratoff}},\ }\bibfield  {title} {\bibinfo {title} {Tunneling between
  superconductors},\ }\href {https://doi.org/10.1103/PhysRevLett.10.486}
  {\bibfield  {journal} {\bibinfo  {journal} {Physical Review Letters}\
  }\textbf {\bibinfo {volume} {10}},\ \bibinfo {pages} {486} (\bibinfo {year}
  {1963})}\BibitemShut {NoStop}%
\bibitem [{\citenamefont {Kulik}\ and\ \citenamefont
  {Omel'yanchuk}(1975)}]{Kulik1975_JosephsonEffect}%
  \BibitemOpen
  \bibfield  {author} {\bibinfo {author} {\bibfnamefont {I.~O.}\ \bibnamefont
  {Kulik}}\ and\ \bibinfo {author} {\bibfnamefont {A.~N.}\ \bibnamefont
  {Omel'yanchuk}},\ }\bibfield  {title} {\bibinfo {title} {Contribution to the
  microscopic theory of the {Josephson} effect in superconducting bridges},\
  }\href {https://www.osti.gov/biblio/4209268} {\bibfield  {journal} {\bibinfo
  {journal} {JETP Letters}\ }\textbf {\bibinfo {volume} {21}},\ \bibinfo
  {pages} {96} (\bibinfo {year} {1975})}\BibitemShut {NoStop}%
\bibitem [{\citenamefont {Kulik}\ and\ \citenamefont
  {Omel'yanchuk}(1977)}]{Kulik1977_Microbridges}%
  \BibitemOpen
  \bibfield  {author} {\bibinfo {author} {\bibfnamefont {I.~O.}\ \bibnamefont
  {Kulik}}\ and\ \bibinfo {author} {\bibfnamefont {A.~N.}\ \bibnamefont
  {Omel'yanchuk}},\ }\bibfield  {title} {\bibinfo {title} {Properties of
  superconducting microbridges in the pure limit},\ }\href
  {https://www.osti.gov/biblio/6927703} {\bibfield  {journal} {\bibinfo
  {journal} {Sov. J. Low Temp. Phys. (Engl. Transl.)}\ }\textbf {\bibinfo
  {volume} {3}},\ \bibinfo {pages} {459} (\bibinfo {year} {1977})}\BibitemShut
  {NoStop}%
\bibitem [{\citenamefont {Likharev}(1979)}]{Likharev1979_WeakLinks}%
  \BibitemOpen
  \bibfield  {author} {\bibinfo {author} {\bibfnamefont {K.~K.}\ \bibnamefont
  {Likharev}},\ }\bibfield  {title} {\bibinfo {title} {Superconducting weak
  links},\ }\href {https://doi.org/10.1103/RevModPhys.51.101} {\bibfield
  {journal} {\bibinfo  {journal} {Reviews of Modern Physics}\ }\textbf
  {\bibinfo {volume} {51}},\ \bibinfo {pages} {101} (\bibinfo {year}
  {1979})}\BibitemShut {NoStop}%
\bibitem [{\citenamefont {Tinkham}(2004)}]{tinkham2004introduction}%
  \BibitemOpen
  \bibfield  {author} {\bibinfo {author} {\bibfnamefont {M.}~\bibnamefont
  {Tinkham}},\ }\href@noop {} {\emph {\bibinfo {title} {Introduction to
  Superconductivity}}},\ \bibinfo {edition} {2nd}\ ed.\ (\bibinfo  {publisher}
  {Dover Publications},\ \bibinfo {year} {2004})\BibitemShut {NoStop}%
\bibitem [{\citenamefont {Jarillo-Herrero}\ \emph {et~al.}(2006)\citenamefont
  {Jarillo-Herrero}, \citenamefont {van Dam},\ and\ \citenamefont
  {Kouwenhoven}}]{Jarillo-Herrero2006_Supercurrent}%
  \BibitemOpen
  \bibfield  {author} {\bibinfo {author} {\bibfnamefont {P.}~\bibnamefont
  {Jarillo-Herrero}}, \bibinfo {author} {\bibfnamefont {J.~A.}\ \bibnamefont
  {van Dam}},\ and\ \bibinfo {author} {\bibfnamefont {L.~P.}\ \bibnamefont
  {Kouwenhoven}},\ }\bibfield  {title} {\bibinfo {title} {Quantum supercurrent
  transistors in carbon nanotubes},\ }\href
  {https://doi.org/10.1038/nature04550} {\bibfield  {journal} {\bibinfo
  {journal} {Nature}\ }\textbf {\bibinfo {volume} {439}},\ \bibinfo {pages}
  {953} (\bibinfo {year} {2006})}\BibitemShut {NoStop}%
\bibitem [{\citenamefont {Joyez}\ \emph {et~al.}(1994)\citenamefont {Joyez},
  \citenamefont {Lafarge}, \citenamefont {Filipe}, \citenamefont {Esteve},\
  and\ \citenamefont {Devoret}}]{1994_ParitySuppression_Devoret}%
  \BibitemOpen
  \bibfield  {author} {\bibinfo {author} {\bibfnamefont {P.}~\bibnamefont
  {Joyez}}, \bibinfo {author} {\bibfnamefont {P.}~\bibnamefont {Lafarge}},
  \bibinfo {author} {\bibfnamefont {A.}~\bibnamefont {Filipe}}, \bibinfo
  {author} {\bibfnamefont {D.}~\bibnamefont {Esteve}},\ and\ \bibinfo {author}
  {\bibfnamefont {M.~H.}\ \bibnamefont {Devoret}},\ }\bibfield  {title}
  {\bibinfo {title} {Observation of parity-induced suppression of
  {Josephson}tunneling in the superconducting single electron transistor},\
  }\href {https://doi.org/10.1103/PhysRevLett.72.2458} {\bibfield  {journal}
  {\bibinfo  {journal} {Physical Review Letters}\ }\textbf {\bibinfo {volume}
  {72}},\ \bibinfo {pages} {2458} (\bibinfo {year} {1994})}\BibitemShut
  {NoStop}%
\bibitem [{\citenamefont {Hertel}\ \emph {et~al.}(2021)\citenamefont {Hertel},
  \citenamefont {Andersen}, \citenamefont {van Zanten}, \citenamefont
  {Eichinger}, \citenamefont {Scarlino}, \citenamefont {Yadav}, \citenamefont
  {Karthik}, \citenamefont {Gronin}, \citenamefont {Gardner}, \citenamefont
  {Manfra}, \citenamefont {Marcus},\ and\ \citenamefont
  {Petersson}}]{Hertel2021_SAG}%
  \BibitemOpen
  \bibfield  {author} {\bibinfo {author} {\bibfnamefont {A.}~\bibnamefont
  {Hertel}}, \bibinfo {author} {\bibfnamefont {L.~O.}\ \bibnamefont
  {Andersen}}, \bibinfo {author} {\bibfnamefont {D.~M.~T.}\ \bibnamefont {van
  Zanten}}, \bibinfo {author} {\bibfnamefont {M.}~\bibnamefont {Eichinger}},
  \bibinfo {author} {\bibfnamefont {P.}~\bibnamefont {Scarlino}}, \bibinfo
  {author} {\bibfnamefont {S.}~\bibnamefont {Yadav}}, \bibinfo {author}
  {\bibfnamefont {J.}~\bibnamefont {Karthik}}, \bibinfo {author} {\bibfnamefont
  {S.}~\bibnamefont {Gronin}}, \bibinfo {author} {\bibfnamefont {G.~C.}\
  \bibnamefont {Gardner}}, \bibinfo {author} {\bibfnamefont {M.~J.}\
  \bibnamefont {Manfra}}, \bibinfo {author} {\bibfnamefont {C.~M.}\
  \bibnamefont {Marcus}},\ and\ \bibinfo {author} {\bibfnamefont {K.~D.}\
  \bibnamefont {Petersson}},\ }\bibfield  {title} {\bibinfo {title} {Electrical
  properties of selective-area-grown superconductor-semiconductor hybrid
  structures on silicon},\ }\href
  {https://doi.org/10.1103/PhysRevApplied.16.044015} {\bibfield  {journal}
  {\bibinfo  {journal} {Physical Review Applied}\ }\textbf {\bibinfo {volume}
  {16}},\ \bibinfo {pages} {044015} (\bibinfo {year} {2021})}\BibitemShut
  {NoStop}%
\bibitem [{\citenamefont {Su}\ \emph {et~al.}(2018)\citenamefont {Su},
  \citenamefont {Zarassi}, \citenamefont {Hsu}, \citenamefont {San-Jose},
  \citenamefont {Prada}, \citenamefont {Aguado}, \citenamefont {Lee},
  \citenamefont {Pendharkar}, \citenamefont {Lee}, \citenamefont {Op~het
  Veld},\ and\ \citenamefont {et~al.}}]{Sergey2018Mirage}%
  \BibitemOpen
  \bibfield  {author} {\bibinfo {author} {\bibfnamefont {Z.}~\bibnamefont
  {Su}}, \bibinfo {author} {\bibfnamefont {A.}~\bibnamefont {Zarassi}},
  \bibinfo {author} {\bibfnamefont {J.-F.}\ \bibnamefont {Hsu}}, \bibinfo
  {author} {\bibfnamefont {P.}~\bibnamefont {San-Jose}}, \bibinfo {author}
  {\bibfnamefont {E.}~\bibnamefont {Prada}}, \bibinfo {author} {\bibfnamefont
  {R.}~\bibnamefont {Aguado}}, \bibinfo {author} {\bibfnamefont {E.~J.~H.}\
  \bibnamefont {Lee}}, \bibinfo {author} {\bibfnamefont {M.}~\bibnamefont
  {Pendharkar}}, \bibinfo {author} {\bibfnamefont {J.~S.}\ \bibnamefont {Lee}},
  \bibinfo {author} {\bibfnamefont {R.~L.~M.}\ \bibnamefont {Op~het Veld}},\
  and\ \bibinfo {author} {\bibnamefont {et~al.}},\ }\bibfield  {title}
  {\bibinfo {title} {Mirage andreev spectra generated by mesoscopic leads in
  nanowire quantum dots},\ }\href
  {https://doi.org/10.1103/PhysRevLett.121.127705} {\bibfield  {journal}
  {\bibinfo  {journal} {Physical Review Letters}\ }\textbf {\bibinfo {volume}
  {121}},\ \bibinfo {pages} {127705} (\bibinfo {year} {2018})}\BibitemShut
  {NoStop}%
\end{thebibliography}%

\clearpage


\renewcommand{\appendixname}{Supplementary Material}
\renewcommand{\thefigure}{S\arabic{figure}} \setcounter{figure}{0}
\renewcommand{\thetable}{S\arabic{table}} \setcounter{table}{0}
\renewcommand{\theequation}{S\arabic{table}} \setcounter{equation}{0}
\renewcommand{\thesection}{S\arabic{section}} \setcounter{section}{0}

\title{Supplementary Information: Engineering Josephson Junctions with Sn-InAs nanowires}

\maketitle

\onecolumngrid

\section{Addendum on Materials}

\subsection{Low magnification images of the annealing temperature series}

For a complete set of images, find the images and their analysis on the online repository linked to the manuscript.

\begin{figure*}[h!] 
  \includegraphics[width=\textwidth]{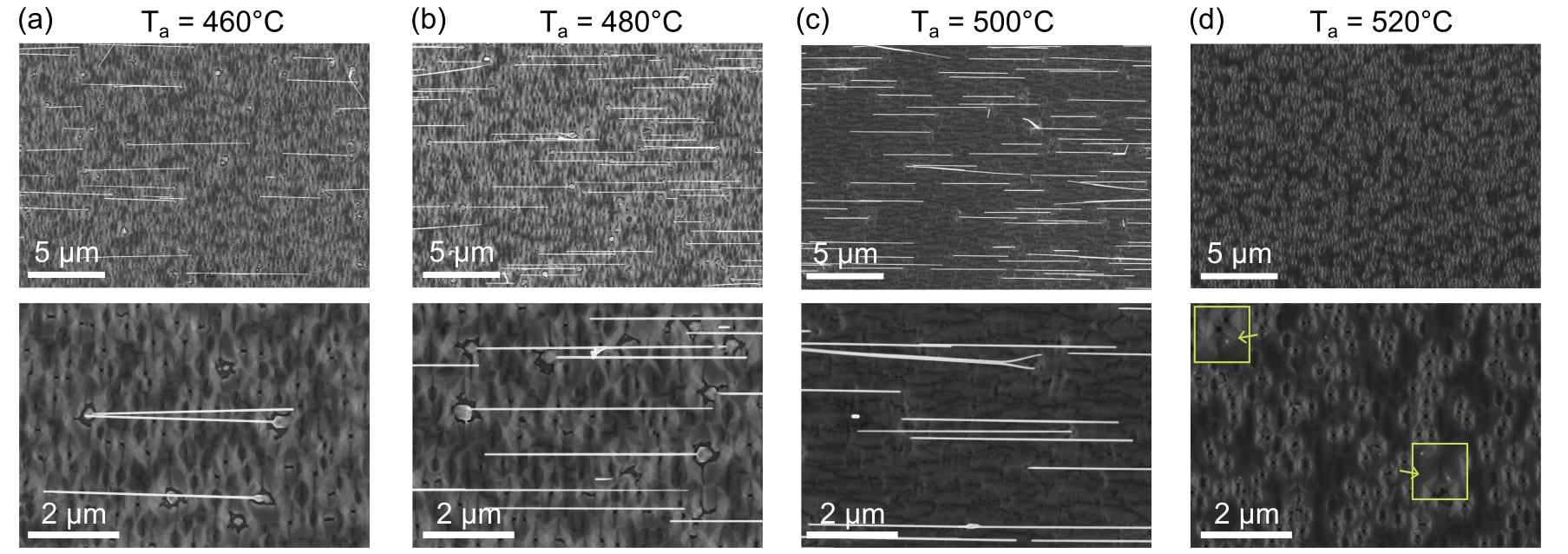}
  \caption{Additional Data for nanowire growth. (a) to (d) Evolution of the nanowire length with annealing temperature (Top view images). Panel (d) shows yellow arrows indicating the position of gold droplets that did not nucleate into nanowires.}
  \label{fig_suppl:Annealing}
\end{figure*}

\newpage
\subsection{Nanowires misalignment after low annealing temperatures}
\begin{figure*}[h!] 
\includegraphics[width=\textwidth]{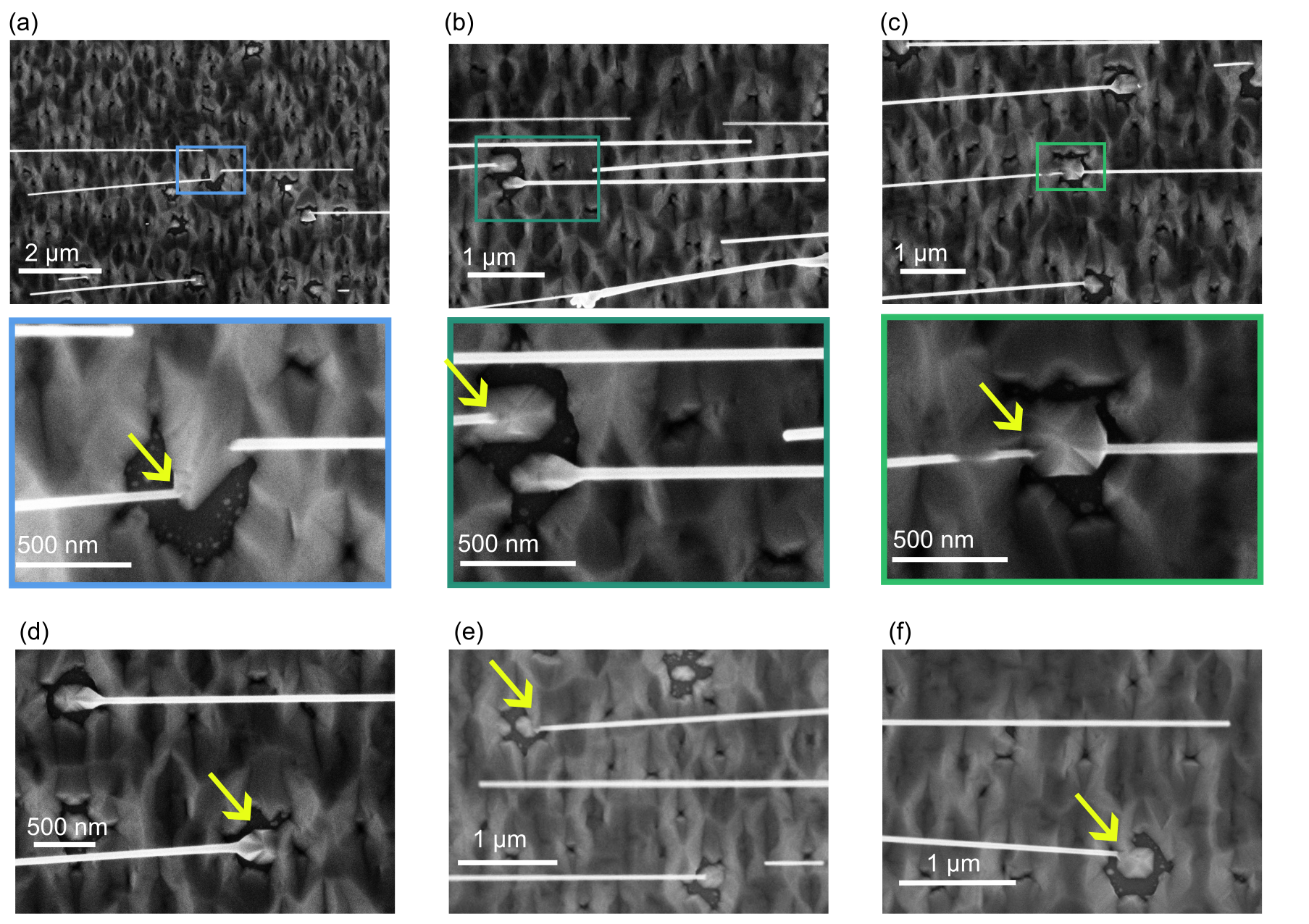}
  \caption{(a) to (f) SEM images of pedestals from which  nanowires grow misaligned after annealing at 460$^{\circ}$C. The yellow arrows indicate the position of the misalignment issue: low definition of the {111}B facet, misalignment of the pedestal with respect to the substrate and spurious 3D growth.}
  \label{fig_suppl:Misoriented}
\end{figure*}

\newpage
\subsection{Initial growth stage of InAs nanowires on (001)InAs}
\begin{figure*}[h!] 
  \includegraphics[width=0.5\textwidth]{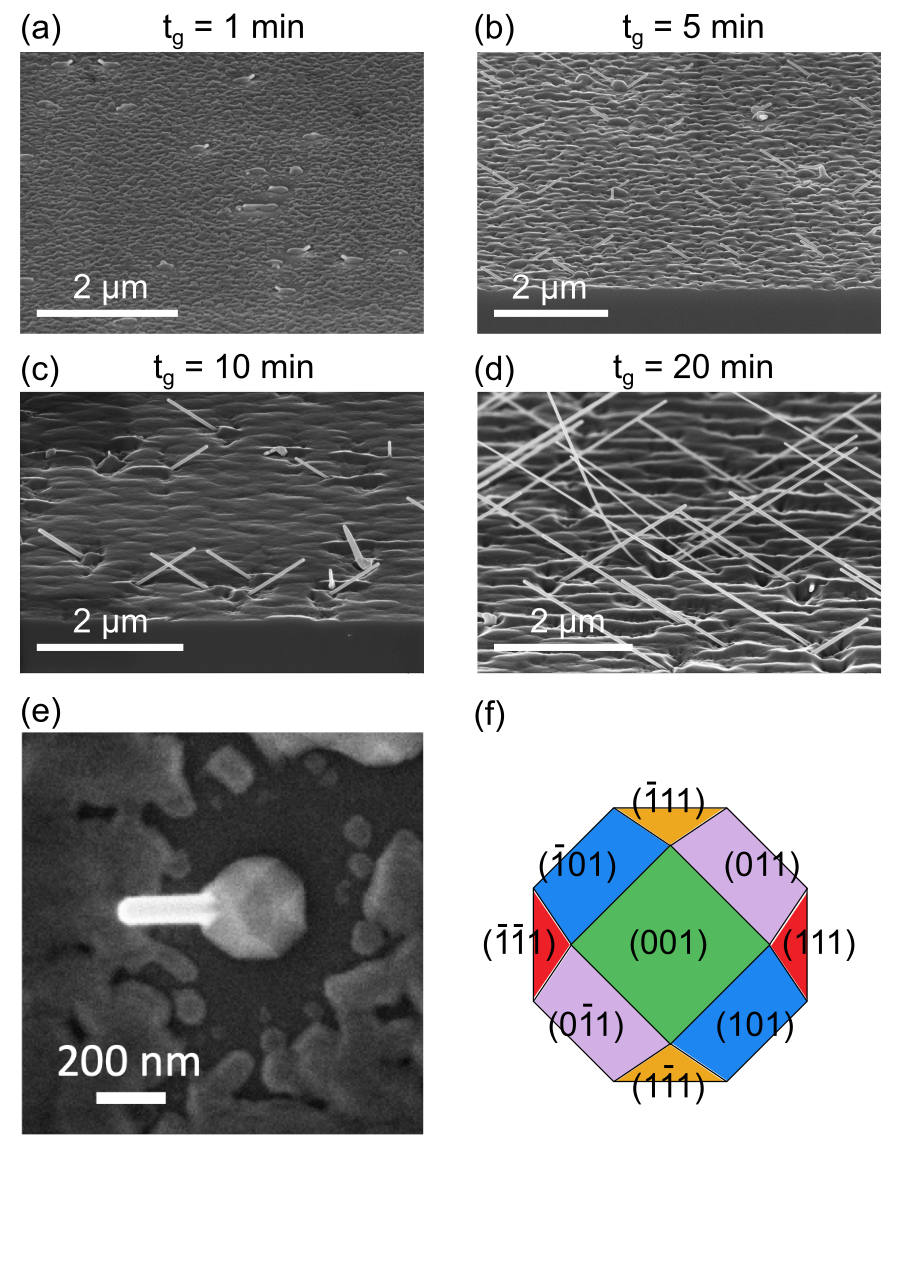}
  \caption{Additional Data for nanowire growth. (a) to (d) Evolution of the nanowire length with growth time (images tilted at 20$^{\circ}$). (e) Top  view image of a nanowire grown from a 50 nm diameter gold catalyst after $t_g$ = 5 min. A faceted mound is clearly visible. (f) Schematic of the facets of the mound with corresponding planes labels.}
  \label{fig_suppl:InitGrowthStage}
\end{figure*}

\newpage
\subsection{Sn grains orientation and lateral dimensions}
\label{SupSec:GrainTEM}

\begin{figure*}[h!] 
  \includegraphics[width=\textwidth]{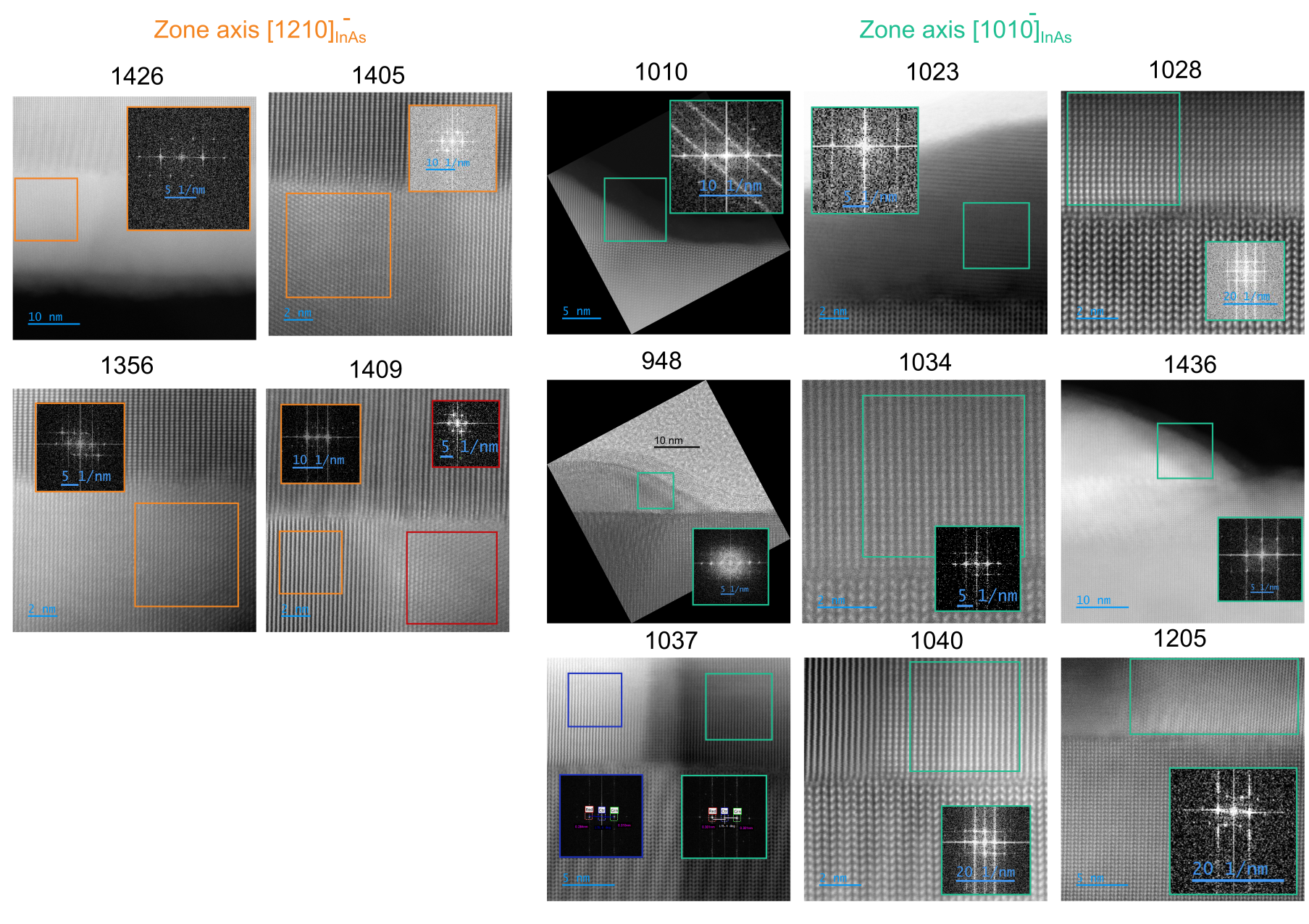}
  \caption{Sn grains analysis. High-resolution TEM images of several Sn grains. First the InAs nanowire is orientation along either the [1210] or [1010] zone axis. Then, the grains with at least one visible crystalline orientation are analyzed using the DiffTools script available online at http://www.dmscripting.com/difftools.html.}
  \label{fig_suppl:Sn_grains}
\end{figure*}

\newpage
\begin{longtable}{|c|c|c|c|c|c|c|c|c|c|}
\hline
\textbf{Grain} & \textbf{Image name} & \textbf{Material} & \textbf{$d_m$ (pm)}& \textbf{ $d_c$ (pm)} & \textbf{Angle ($^{\circ}$)} & \textbf{Direction} & \textbf{Direction} & \textbf{$\beta$ Sn} & \textbf{Comment} \\
&&&&&&Sn&InAs&&\\
\hline
\endfirsthead
        1     & 1356   & Sn & 295  & 287  & 60  & [200]   & [0001] & yes  & slightly off [200] \\
              &        &   & 279  & 272  &     & [101]   &        &     & only 1 direction aligned \\\hline
        2     & 1405   & Sn left & 278  & 270  & 65  & [101]   &        & yes    & slightly off [200] \\
              &        &   & 292  & 284  &     & [200]   & [0001] &       & only 1 direction aligned \\\hline
        3     & 1409   & Sn left & 304  & 296  &     & [200]   & [0001] & yes  & only 1 direction aligned \\\hline
        4     & 1409   & Sn right & 290  & 282  &     & [101]   &        & yes    & slightly off \\
              &        &   & 300  & 292  &     & [200]   & [0001] &       & only 1 direction aligned \\\hline
        5     & 1426   & Sn & 298  & 290  & 60  & [101]   &        & yes    & only 1 direction aligned \\
              &        &   & 300  & 292  &     & [200]   & [0001] &       &  \\\hline
        6     &        & Sn & 284  & 276  & 90  & [020]   & [10-10] &       & slightly off \\
              &        &   & 307  & 298  &     & [200]   & [0001] & yes   & only one direction visible \\\hline
        7     & 1037   & Sn left & 298  & 289  &     & [020]   & [0001] & yes   & only one direction visible \\
              &        &           &     &     &        &        &      &  & slightly off \\\hline
        8     & 1037   & Sn right & 298  & 290  & 90  & [020]   & [10-10] &       &  \\
              &        &   & 285  & 277  &     & [200]   & [0001] & yes   &  \\\hline
        9     & 1023   & Sn right & 293  & 285  & 90  & [200]   & [0001] & yes   & off \\
              &        &   & 165  & 160  &     & [031]   & [10-10] &       &  \\\hline
        10    & 1010   & Sn & 284  & 276  &     & [101]   & [0001] &       &  \\
              &        &   & 284  & 276  & 90  & [011]   & [10-10] & yes   &  \\\hline
        11    & 1028   & Sn & 274  & 266  &     & [101]   & [0001] &       &  \\
              &        &   & 292  & 284  & 90  & [020]   & [10-10] & yes   &  \\\hline
        12    & 1040   & Sn & 286  & 278  &     & [101]   & [0001] &       &  \\
              &        &   & 286  & 278  & 92  & [011]   & [10-10] & yes   &  \\\hline
        13    & 1436   & Sn & 290  & 282  &     & [200]   & [0001] &       &  \\
              &        &   & 289  & 281  & 65  & [101]   &        & yes    &   \\\hline
        14    & 1205   & Sn & 297  & 288  & 60  & [200]   & [0001] &       & only 1 direction aligned \\
              &        &   & 286  & 278  &     & [101]   &        & yes    &  \\\hline
        15    & 1034   & Sn & 283  & 275  & 90  & [011]   & [10-10] &  yes     &  \\
              &        &   & 299  & 291  &     & [200]   & [0001] &       &  \\
\hline

\caption{List of Sn grains studied and their crystalline orientations relative to the crystalline orientation of the InAs nanowire. The interplanar distance of Sn, $d_m$, is measured from the FFT of the TEM image. The corrected interplanar distance, $d_c$, is obtained by scaling $d_m$ using the ratio of the theoretical interplanar distance of InAs, 355 pm, to the measured interplanar distance $d_{[0002]}$ of InAs, following the relation $d_c = d_m \times \frac{355}{d_{[0002]}}$.}

\label{sup_info:grain_orientation}
\end{longtable}

\subsection{Moiré patterns}
Moiré patterns are formed by the superposition of the Sn grains and InAs nanowire lattices. The software used for the simulation can be found online https://unh2d.weebly.com/moire-pattern-simulator.html.

\begin{figure*}[h!] 
  \includegraphics[width=\textwidth]{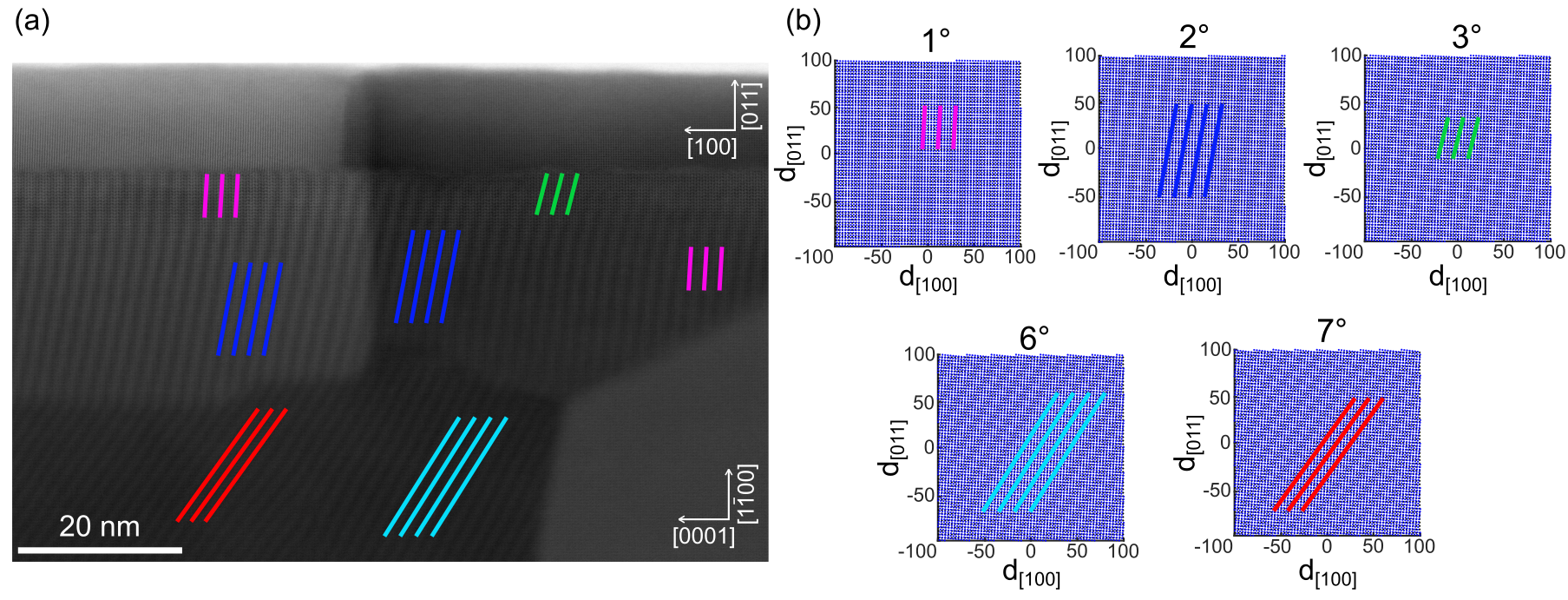}
  \caption{Moiré pattern analysis. (a) HRTEM-HAADF image of a InAs nanowire overlapped with epitaxial Sn grains. (b) Calculated Moiré patterns with different angle of rotation between the planar orientations indicated on the image .}
  \label{fig_suppl:Moire}
\end{figure*}

\section{Addendum on Transport Data}

\subsection{Accounting for background resistance} \label{SupSec:TwoTermResisance}

\paragraph{Ibias experiments}: Since the measurements are two-terminal, we expect a linear current-voltage response even in the superconducting region due to line resistance. We correct it, by subtracting a linear fit to the response before the supercurrent switches to normal state, so that the processed data looks flat in the superconducting region. The measured voltages so corrected, thus correspond to voltage appearing across the nanowire junction. Any further processing to the data is performed after this stage. This is however not possible, in cases when the device is pinched-off (at negative gate voltages) or supercurrent is not present (at higher magnetic fields), the background resistance for that dataset is corrected using the mean differential resistance obtained in the superconducting portions of the raw dataset.

\paragraph{Vbias experiments}: To account for the fridge line resistances in Vbias experiments, we consider the mean of the point resistances seen in the saturation regime of the nanowire obtained from gate sweeps at fixed 10mV external bias. In the saturation regime, assuming the nanowire is fully open, the voltage drop across the nanowire is minimal and the external voltage drops mostly in the measurement setup resistances. All data and $V_{bias}$ axes in these Vbias measurements are normalised according to this value.

\subsection{Inferring induced superconducting gap} \label{SupSec:IndGap}
To infer the induced superconducting gap, we consider the processed data from Vbias experiments near pinch off. At gate voltages where only peaks corresponding to $4\Delta$ resonance are visible, we approximate $\Delta_{ind}$ as the quarter of spacing between $V_{bias}$ where those peaks appear.


\begin{figure*} 
  \includegraphics[width=\textwidth]{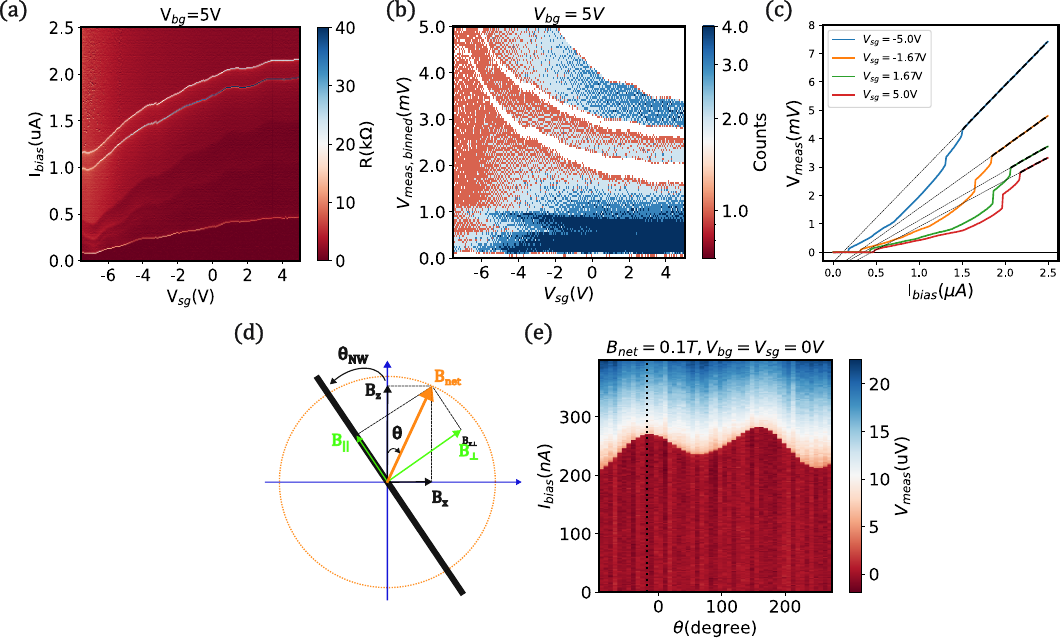}
  \caption{Additional Data for Device A in the main text. (a) Differential resistance maps at different $V_{sg}$ with $I_{bias}$ swept to higher values for extraction of various parmaeters $I_{sw}$, $R_n$, $I_{excess}$. (b) Binning data in (a) to show some resonances correspond to MARs. (c) Illustrating the procedure for extracting $R^*$. (d) Magnetic fields $B_x$ and $B_z$ amounting to net magnetic field $B_{net}$, experienced as $B_{||}$ and $B_{\perp}$ by the nanowire, are varied in a way that rotates $B_{net}$ in the plane of the device chip. When $\theta = \theta_NW$, $B_{net} = B_{||}$ and $B_{||} = B_x /sin(\theta)$. (e) Modulation of $I_{sw}$ seen in $I_{bias}$ sweeps at different rotation angles. Location of maximum $I_{sw}$ marked with dotted black vertical line implies $\theta = \theta_NW$. (f) Normalised excess current as a function of gate.}
  \label{fig_suppl:AdditionalData_DeviceA}
\end{figure*}


\subsection{Multiple Andreev Reflections in $I_{bias}$ Data} \label{SupSec:MARS}
Fig.~\ref{fig_suppl:AdditionalData_DeviceA}a shows differential resistance at different side gate voltages similar to main text at $V_{bg} =-7.5V$ with bias currents sweeped to larger values. Switching currents $I_{sw}$ are again identified as the lowest-current-valued peaks. At currents slightly above $I_{sw}$, some resonant bands separated by shallow dV/dI peaks appear. For every gate, we plot a histogram by binning the voltage across the device and counting the number of applied current bias points as shown in Fig.\ref{fig_suppl:AdditionalData_DeviceA}b. Here, clusters of higher counts corresponding to higher conductance appear at lower voltage bin values and higher gate values. At negative gate voltages, these clusters split and become narrower at constant-voltage bin-values. These features below 1-2mV are similar to resonances obtained in conductance measurements in Vbias experiments as shown in main text c.f. Fig.~\ref{fig:Vbias} and can be attributed to MARS.

\subsection{More on $I_{sw}R_{N}$} \label{SupSec:IswRn}
Here, we explain our procedure of obtaining $I_{sw}R_n$ products. After flattening the superconducting region in the raw data from Ibias experiments, the current-voltage characteristics looks is as shown in Fig.~\ref{fig_suppl:AdditionalData_DeviceA}c for a few side gate values at $V_{bg}= 5V$. There are multiple switches seen in these characteristics. The first switch is the switching current $I_{sw}$. Other switches below $2\Delta/e$ voltage across the device correspond to MARS. We obtain $R^*$ as the slope from linear fits to the segments above the highest switches. These fits extrapolate closest to 0 and may or may not correspond to $R_N$ at all gate settings.

\subsection{Orienting Magnetic Field Along Nanowire} \label{SupSec:OrientB}
To orient the magnetic field along the axis of the nanowire, we first align our device chip plane with the geometric plane defined by the two axes, known apriori, of two magnets X and Z that are available in our measurement setup. We rotate a small and fixed magnitude of magnetic field $B_{net}$ in the plane of device chip as shown in Fig.~\ref{fig_suppl:AdditionalData_DeviceA}d. We track switching currents at different rotation angles $\theta$ with respect to the two independent magnet axes X and Z. Superconductivity is suppressed the most in orientations perpendicular to the nanowire. The orientations ($\theta = \theta_{NW} \pm 0 ~ or ~ 180 ^{\circ}$) with largest switching currents thus correspond to axial direction of the nanowire, marked as dotted vertical line in Fig.~\ref{fig_suppl:AdditionalData_DeviceA}e. The so obtained angle agrees with the orientation of the nanowire with respect to the chip which in turn is oriented with respect to the PCB (mounting the chip on PCB by hand can also add/subtract some degrees) placed in the frame defined by the two magnets. With this method we estimate a misalignment error of around $\pm 7 ^{\circ}$ set by the resolution of angular scans. 

\subsection{Observation of skew}
For Device A, in magnetic field sweeps in Ibias experiments we observe a skew while tracking switching currents vs magnetic fields in certain gate combinations. In this experiment the current is only swept up in magnitude in either directions and is different from hysteresis obtained in underdamped junction. This skew could point towards anomalous josephson effects in the nanowire junction \cite{zhang2022evidence}. 


\begin{figure*}
  \includegraphics[width=\textwidth]{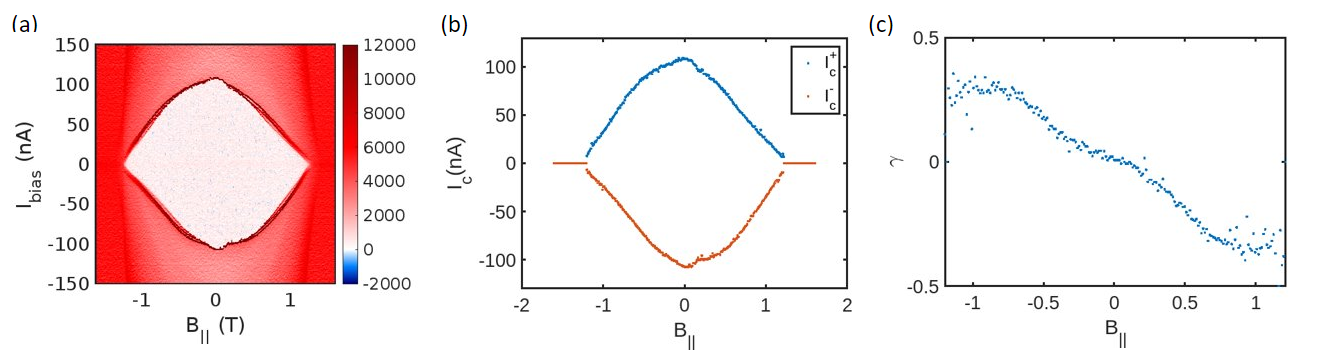}(
  \caption{Skew seen in Device A. (a) Differential conductance at $V_{bg}= -7.5V, V_{sg} = 2.5V$ as a functiong of $B_{||}$ when current is swept up in magnitude in both positive and negative directions (b) extracted switching currents in positive ($I_{sw}^+$) and negative directions ($I_{sw}^-$) (c) Skew parameter $\gamma = \frac{|I_{sw}^+|-|I_{sw}^-|}{ \left( \frac{|I_{sw}^+|+|I_{sw}^-|}{2} \right) }$ }
\end{figure*}

\section{Data from other junction devices}
A summary of extracted parameters from measured SNS devices across the two cooldowns is summarised in Table.~\ref{table:summarySNS}

\begin{table}[h]
\caption{\label{table:summarySNS}Summary of extracted parameters}
\begin{ruledtabular}
\begin{tabular}{ccccc}
Device (internal label), Cooldown \# & max $I_{sw}(nA)$ & $\Delta_{ind}(\mu eV)$ & Critical $B_{||}(T)$ & $I_{sw}R^* (\mu V)$ \\
\hline
A (1.24), 1                 & 475   & 630 & 1.2 & 400-790 \\
B (1.33), 1                 & 146   & 580 & 1.5 & 200-390 \\
C (4.24\textunderscore2), 2 & 185   & 580 & -   & 280-450 \\
D (3.32), 1                 & 360   & -   & 1.7 & 200-1020 \\
E (1.22), 2                 & 278 & 580 & -   & 280-560 \\
\end{tabular}
\end{ruledtabular}
\end{table}

\begin{figure*}
  \includegraphics[width=\textwidth]{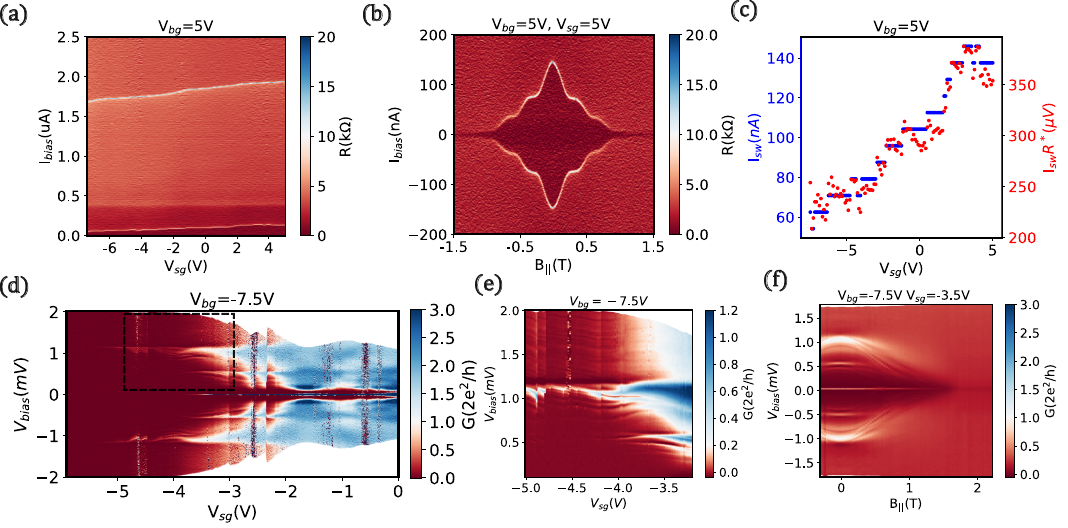}
  \caption{Device B measurements. (a) Differential resistance maps at different $V_{sg}$ with $I_{bias}$ at $V_{bg} = 5V$. (b) Differnetial resistance as a function of $I_{bias}$ and $B_{||}$ at $V_{bg} = V_{sg} = 5V$. (c) Extracted $I_{sw}$ and $I_{sw}R^*$ products from (b) on left and right axes respectively at different gates. (d) Differential Conductance near pinch-off as a function of $V_{bias}$ and $V_{sg}$ at $V_{bg}=-7.5V$. MARS are noted as finite bias peaks and supercurrent as zero bias peak. (e) Zoomed-in conductance measurements near finite bias peaks in (d). We see extra resonant fringes near the prominent MARs peaks and could be a result of non-trivial density of states of in superconducting leads \cite{Sergey2018Mirage}. (f) Differential conductance as a function of $V_{bias}$ and $B_{||}$ shows the evolution of sub-gap features with magnetic field. The superconducting gap closes and the zero-bias current dies at about 1.5\unit{T}.}
\end{figure*}


\begin{figure*}
  \includegraphics[width=\textwidth]{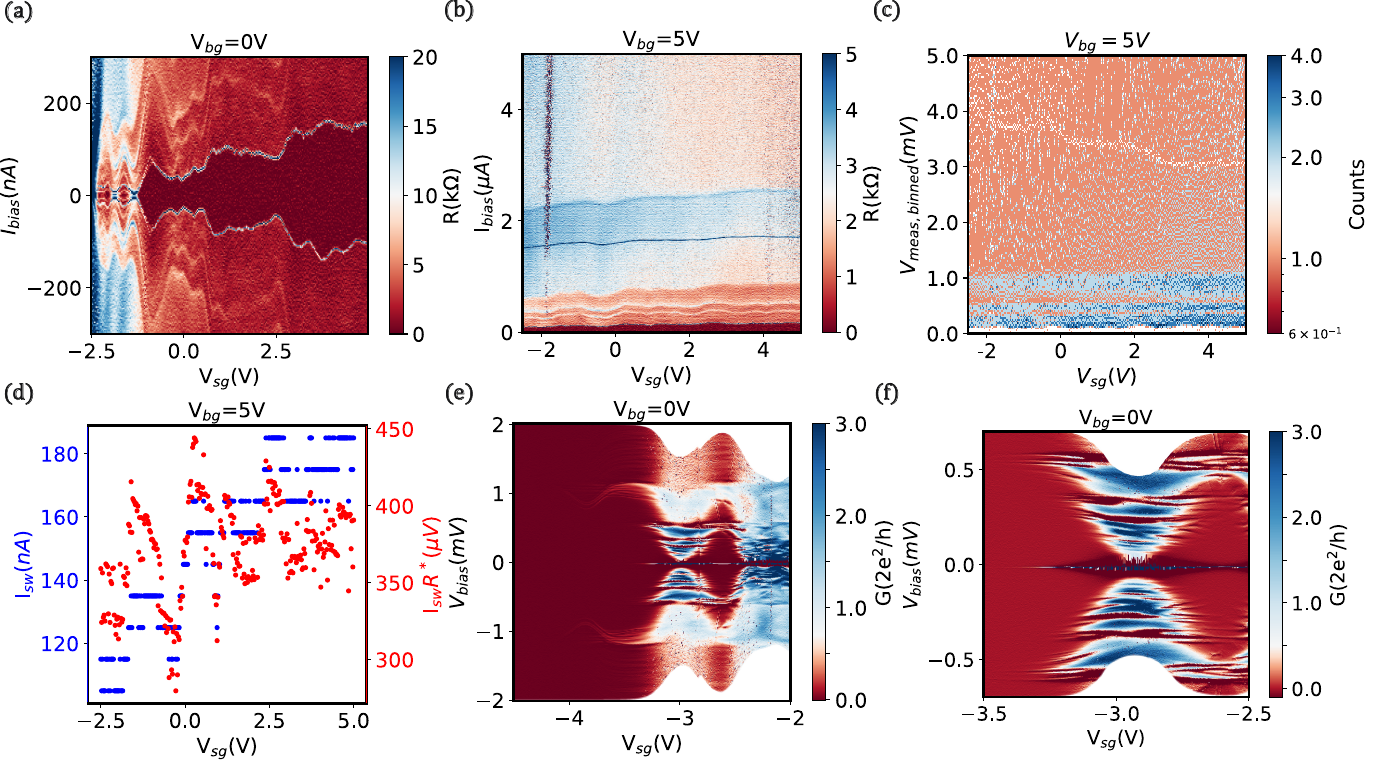}
  \caption{Device C measurements. (a) Differential resistance maps at different $V_{sg}$ with $I_{bias}$ at $V_{bg} = 0V$. (b) Differnetial resistance as a function of higher bias currents $I_{bias}$ and $V_{sg}$ at $V_{bg} = 5V$. (c) Binning data in (a) to show some resonances correspond to MARs. (d) Extracted $I_{sw}$ and $I_{sw}R^*$ products from (b) on left and right axes respectively at different gates. (e) Differential Conductance near pinch-off as a function of $V_{bias}$ and $V_{sg}$ at $V_{bg}=-0V$. MARs are noted as finite bias peaks and supercurrent as zero bias peak. (f) Zoomed-in conductance measurements near finite bias peaks in (e). MARs upto order 5 are discernible. In addition we see extra resonant fringes similar to Device B near the prominent MARs peaks and could be a result of non-trivial density of states of in superconducting leads}
\end{figure*}
\begin{figure*}
  \includegraphics[width=0.7\textwidth]{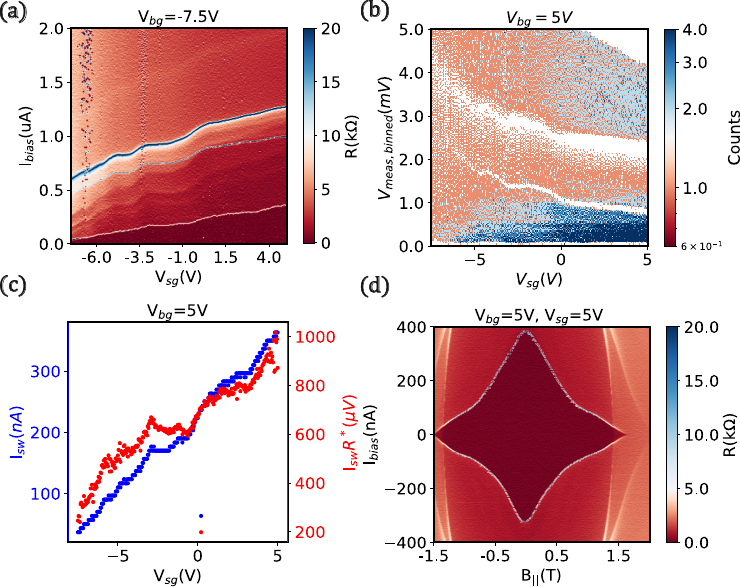}
  \caption{Device D measurements. (a) Differential resistance maps at different $V_{sg}$ with $I_{bias}$ at $V_{bg}=5V$. (b) MARS features seen when the data in (a) is histogrammed as explained in \ref{SupSec:MARS}. (c) Extracted $I_{sw}$ and $I_{sw}R^*$ products from (a) on left and right axes respectively. (d) Differnetial resistance as a function of $I_{bias}$ and $B_{||}$ at $V_{bg} = V_{sg} = 5V$. Supercurrents survive upto critical $B_{||}~1.7T$}
\end{figure*}

\begin{figure*}
  \includegraphics[width=\textwidth]{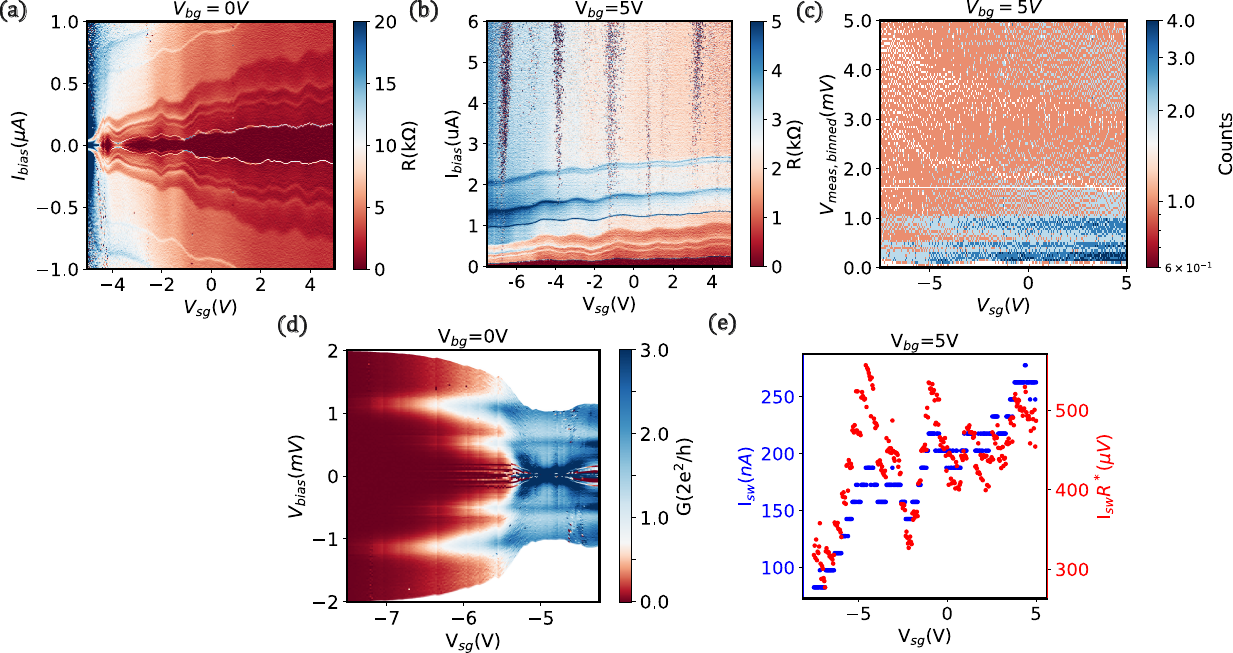}
  \caption{Device E measurements. (a) Differential resistance maps at different $V_{sg}$ with $I_{bias}$ at $V_{bg}=0V$. (b) Differential resistance maps at different $V_{sg}$ with higher $I_{bias}$ at $V_{bg}=5V$. (c) MARS features seen when the data in (a) is histogrammed as explained in \ref{SupSec:MARS}. (d) Extracted $I_{sw}$ and $I_{sw}R^*$ products from (b) on left and right axes respectively.} 
\end{figure*}

\section{Additional Data on a Shell type device}


\begin{figure*}
  \includegraphics[width=0.9\textwidth]{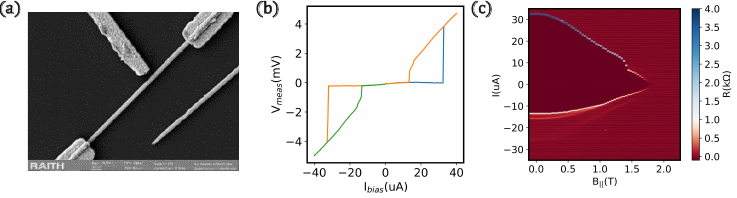}
  \caption{Device F measurements. (a) SEM image of Device F with slightly modulated but un-interrupted Sn shell on InAs nanowire. The continuity of the Sn shell was discerned from gate measurements where no gate control was observed accompanied by high switching currents up to 32nA shown in (b). (c) Differential resistance at zero magnetic field and floating gate.}
\end{figure*}

\end{document}